\documentclass[fleqn,usenatbib]{mnras}

\usepackage{newtxtext,newtxmath}
\usepackage[T1]{fontenc}

\DeclareRobustCommand{\VAN}[3]{#2}
\let\VANthebibliography\thebibliography
\def\thebibliography{\DeclareRobustCommand{\VAN}[3]{##3}\VANthebibliography}

\usepackage{graphicx}	% Including figure files
\usepackage{amsmath}	% Advanced maths commands
\usepackage{braket}
\usepackage{siunitx}
\title[New scaling relations for subgiant stars]{New scaling relations from MESA models for mass and radius of subgiant stars: application to \emph{Kepler} targets}
\author[T. Çakır Alsaç, M. Yıldız and S. Örtel] {T. Çakır Alsaç,$^{1,2}$\thanks{E-mail: tubanurcakir97@gmail.com}\thanks{ORCID: 0000-0002-2538-0105}
M. Yıldız,$^{2}$\thanks{ORCID: 0000-0002-7772-7641},
Sibel Örtel$^{1,2}$\thanks{ORCID: 0000-0001-5759-7790}
\\
$^{1}$Department of Astronomy and Space Sciences, Graduate School of Natural and Applied Sciences, Ege
University, İzmir, Turkey\\
$^{2}$Department of Astronomy and Space Sciences, Faculty of Science, Ege
University, İzmir, Turkey\\
}

\date{Accepted XXX. Received YYY; in original form ZZZ}

\pubyear{2015}

\begin{document}
\label{firstpage}
\pagerange{\pageref{firstpage}--\pageref{lastpage}}
\maketitle

\begin{abstract}
The precise determination of the fundamental parameters of stars is crucial for understanding stellar structure and evolution. In this regard, the asteroseismic parameters of solar-like oscillating stars obtained from space telescopes are very useful for determining the mass ($M$) and radius ($R$) of single stars using classical scaling relations. These relations still need to be improved. 
In this study, we develop alternative scaling relations based on reference frequencies ($\nu_{\rm min}$) due to helium ionization zone glitches, specifically for subgiant (SG) stars. We compile main-sequence and SG {\small{MESA}} models from the literature and derive new scaling relations for stellar $R$ and surface gravity ($g$) of evolved stars. The expressions for $R$ and $g$ are obtained simultaneously as functions of the large frequency separation, $\nu_{\rm min}$, and metallicity. These new relations allow $M$ and $R$ to be reliably determined only from observational parameters, without the need for detailed stellar modelling. 
The resulting $\nu_{\rm min}$-based relations are applied to a sample of 31 SG stars observed by the \emph{Kepler} mission, with masses in the range 0.85–1.74 M$_\odot$ and radii between 1.03 and 2.82 R$_\odot$. Using simultaneous solutions, we estimate new mass and radius ranges of 0.89–1.62 M$_\odot$ and 1.04–2.95 R$_\odot$, respectively.
The results demonstrate significant improvements in the determination of $M$ and $R$ for evolved \emph{Kepler} target stars, highlighting the potential of the new scaling relations in asteroseismic analyses.
\end{abstract}


%%%%%%%%%%%%%%%%%%%%%%%%%%%%%%%%%%%%%%%%%%%%%%%%%%

%%%%%%%%%%%%%%%%% BODY OF PAPER %%%%%%%%%%%%%%%%%%
\section{Introduction}
Stars are celestial bodies that generate energy through nuclear reactions in their cores, transport this energy from the interior to the surface, and radiate it into space. Understanding the interior structure and evolution of these objects is one of the main goals of astrophysics. 
%Therefore, it is very important to determine the fundamental parameters of the star: mass ($M$), radius ($R$), luminosity ($L$), age ($t$), helium abundance ($Y$), and heavy element abundance ($Z$).  
Therefore, it is of great importance to determine the stellar mass ($M$), radius ($R$), luminosity ($L$), age ($t$), helium abundance ($Y$), and heavy-element abundance ($Z$). Asteroseismology provides comprehensive information about the deep stellar interior {\citep{2021RvMP...93a5001A}}. One of its main purposes is to determine the fundamental parameters, mass and radius, using the oscillation frequencies of stars.
In this direction, CoRoT \citep[][]{baglin2006}, \emph{Kepler} \citep[][]{borucki2010}, and TESS \citep[][]{ricker2015} space telescopes were launched with the expectation that many planets, in particular Earth-sized and smaller planets in the habitable zone of other stars, would be discovered. These space missions have greatly contributed to the development of asteroseismology. High-quality photometric data have been obtained with more precise and comprehensive measurements. \emph{Kepler}'s success with its accuracy in seismic data is still at the forefront.

%Determining the fundamental parameters of stars with high precision is of great importance. 
For oscillating stars, analysis of oscillation frequencies allows the determination of asteroseismic parameters such as the mean value of the large separation between oscillation frequencies ($\braket{\Delta\nu}$), the frequency of maximum amplitude ($\nu_{\rm max}$), and the mean value of the small separation between oscillation frequencies ($\langle \delta\nu_{02} \rangle$). These seismic parameters enable accurate estimates of the stellar $M$ and $R$ through asteroseismic scaling relations. The basic form of the scaling relations was first proposed by \citet{1995A&A...293...87K}, who related $\nu_{\rm max}$ and $\braket{\Delta\nu}$ to the star's mass and radius.  $\braket{\Delta\nu}$ is related to the square root of the mean stellar density {\citep[$\Delta\nu \propto \sqrt{\bar{\rho}}$;][]{1986ApJ...306L..37U}}  and shifts to lower frequency values as the star evolves. $\nu_{\rm max}$ can be approximated by the relation $\nu_{\rm max} \propto g /\sqrt{T_{\rm eff}}$ {\citep{1991ApJ...368..599B}}, where ${T_{\rm eff}}$ is the effective temperature and $g$ is the surface gravity. 
{Using the observational $\braket{\Delta\nu}$ and $\nu_{\rm max}$ together with $T_{\rm eff}$, the stellar $M$ and $R$ can be written as \citep[e.g. see][]{2009MNRAS.400L..80S,2010A&A...509A..77K,2011Sci...332..213C}:}

%The fundamental parameters of stars can be derived using asteroseismology through scaling relations based on their oscillation frequencies. The basic form of the scaling relations was first proposed by \citet{1995A&A...293...87K}, who related the frequency of maximum amplitude ($\nu_{\rm max}$) and the mean value of the large separation between oscillation frequencies ($\braket{\Delta\nu}$) to the star's mass and radius.
%\textbf{In addition, the determination of the mean value of small frequency separation ($\langle \delta\nu_{02} \rangle$) provides important information for understanding stellar structure and evolution. $\langle \delta\nu_{02} \rangle$ is particularly sensitive to main-sequence (MS) evolution. Determination of seismic parameters such as $\nu_{\rm max}$ and $\langle \Delta\nu \rangle$ is essential for estimating the stellar mass and radius.} $\braket{\Delta\nu}$ is related to the square root of the mean stellar density ($\Delta\nu \propto \sqrt{\bar{\rho}}$) and shifts to lower frequency values as the star evolves. $\nu_{\rm max}$ can be approximated by the relation $\nu_{\rm max} \propto g /\sqrt{T_{\rm eff}}$, where ${T_{\rm eff}}$ is the effective temperature and $g$ is the surface gravity. \textbf{Using the observational $\braket{\Delta\nu}$ and $\nu_{\rm max}$ together with $T_{\rm eff}$, the stellar $M$ and $R$ can be written as \citep[e.g., see][]{2009MNRAS.400L..80S,2010A&A...509A..77K,2011Sci...332..213C}:}
\begin{equation}
{M}/{M_{\odot}}=\frac{(\nu_{\rm max}/\nu_{\rm max\odot})^3}{(\braket{\Delta\nu}/\braket{\Delta\nu}_{\odot})^4}{(T_{\rm eff}/T_{\rm eff\odot})}^{3/2},
    \label{eq:sca_rel_M}
\end{equation}
\begin{equation}
{R}/{R_{\odot}}=\frac{(\nu_{\rm max}/\nu_{\rm max\odot})}{(\braket{\Delta\nu}/\braket{\Delta\nu}_{\odot})^2}{(T_{\rm eff}/T_{\rm eff\odot})}^{1/2}.
    \label{eq:sca_rel_R}
\end{equation} 
Here, $\nu_{{\rm max}\odot}$, $\Delta\nu_{\odot}$, and $T_{{\rm eff}\odot}$ denote the solar reference values (see Table \ref{sismik_gunes_parametreleri}).
Classical scaling relations allow the $M$ and $R$ of stars exhibiting solar-like oscillations to be determined with high precision; however, these relations may show limited accuracy, mainly due to uncertainties in $\nu_{\rm max}$ and structural differences among stars \citep[e.g.][]{2017ApJ...843...11V, 2016MNRAS.460.4277G}.
%Although the classical scaling relations allow the stellar $M$ and $R$ to be determined with high precision for MS stars exhibiting solar-like oscillations, the same agreement does not apply to stars in the post-MS evolutionary phase. These relations may exhibit limited accuracy, mainly due to uncertainties in $\nu_{\rm max}$ and structural differences among stars \citep[e.g.][]{2017ApJ...843...11V, 2016MNRAS.460.4277G}. 
To overcome these limitations, the scaling relations have been revised by several studies with different calibrations and correction terms \citep{2011ApJ...743..161W, 2016ApJ...822...15S, 2016MNRAS.462.1577Y}.
%As shown in equations (\ref{eq:sca_rel_M}) and (\ref{eq:sca_rel_R}), $M$ and $R$ depend on the seismic parameters $\nu_{\rm max}$ and $\Delta\nu$. However, the observational uncertainties associated with $\nu_{\rm max}$ are generally larger than those of $\Delta\nu$. This results in increased uncertainties in the derived stellar parameters, necessitating the development of scaling relations. Therefore, in this study, we propose new scaling relations based on reference frequencies ($\nu_{\rm min0}$, $\nu_{\rm min1}$). 
{As a star evolves, structural changes in its interior directly affect the oscillation frequencies; therefore, calibrations based solely on main-sequence (MS) models may be insufficient for SG stars. Therefore, scaling relations should be calibrated to also include SG stars.}
%In addition, scaling relations should be calibrated to include subgiant (SG) stars in particular, as stars in this evolutionary phase exhibit significant structural and seismic differences compared to MS stars. Since changes in the internal structure directly affect the oscillation frequencies, calibrations based only on MS models are not sufficient for SG stars.

%The development of asteroseismology provides very important information about the interior structure and evolution of solar-like oscillating stars. 
The detected oscillation frequencies are spaced very regularly in solar-like oscillating stars. 
The He II ionization zone, just below the surface, influences the oscillation frequencies \citep[e.g., see][]{1994MNRAS.269..475P,1998MNRAS.295..344P} and causes small glitches at regular intervals. These glitches give rise to new reference frequencies $\nu_{\rm \min}$, which were first identified by \cite{2014MNRAS.441.2148Y}. {Reference frequencies can be determined by plotting a graph of $\Delta\nu$ - $\nu$, where the dips (minima) seen on this graph correspond to the reference frequencies (see Fig. \ref{fig:ref_req_8645}).} {The minimum corresponding to the deepest slope is called $\nu_{\rm min1}$, the minimum at a higher frequency than this point is called $\nu_{\rm min0}$, and the minimum at a lower frequency is called $\nu_{\rm min2}$.} 
$\nu_{\rm min}$ is sensitive to fundamental stellar parameters and internal structural features.  
For example, as the mass increases, the frequency values decrease, and the minima shift towards lower frequencies.
As the star evolves, the frequency values decrease, and while $\nu_{\rm \min2}$ disappears from the observable frequency range, $\nu_{\rm \min1}$ continues to show a deep minimum.

Furthermore, the variations of $\nu_{\rm min}$ and $\nu_{\rm \max}$ with stellar evolution are similar \citep{2014MNRAS.441.2148Y}; therefore, if $\nu_{\rm min}$ can be detected in evolved stars, the $M$ and $R$ can be determined. These minima, unlike parameters such as $\Delta\nu$ and $\nu_{\rm \max}$, which indicate general properties of the entire star, are sensitive to sudden changes in the speed of sound in a certain layer within the star (especially the He II ionization zone) and reflect the physical properties of that layer.
%\textbf{Although the sensitivity of the reference frequencies arises from a localized layer, the location and physical properties of this layer are strongly determined by the global structure of the star. Since He II ionization occurs at a certain temperature, the location of this zone within the star is determined by the temperature profile, which depends primarily on the stellar mass and radius. As the stellar mass and radius change, the temperature profile also changes accordingly. Consequently, the acoustic signature of the He II ionization zone carries indirect but reliable information about global stellar parameters.}
{Although these frequencies are sensitive to a specific physical layer, they can be directly determined from both observational and model oscillation frequencies, similar to $\Delta\nu$. In contrast, $\nu_{\rm \max}$ in models can only be derived from scaling relations. This direct determinability makes the reference frequencies more reliable for estimating global stellar properties.}
Therefore, when calculating $M$ and $R$, reference frequencies can be used (preferably) instead of $\nu_{\rm \max}$, whose uncertainty is relatively high.
{Together with metallicity ($Z$), these quantities ($\Delta\nu$ and $\nu_{\rm min}$) yield important information on the stellar interior structure and evolutionary state \citep{2025MNRAS.538..844O}. Therefore, accurate determination of these parameters is essential for a consistent interpretation of both classical and asteroseismic results.}

%\citet{2019MNRAS.489.1753Y} used reference frequencies in scaling relations to determine the fundamental properties ($T_{\rm eff}$, $M$ and $R$) of MS stars. The use of scaling relations based on reference frequencies has led to significant improvements over classical scaling relations. 

%In this study, we develop similar new asteroseismic methods. The reliability of such methods depends on the accurate and precise determination of the asteroseismic parameters used.

%In particular, we calculate the stellar mass $M$ and radius $R$ using reference frequencies, due to the relatively high uncertainty of $\nu_{\rm max}$ in classical scaling relations compared to other seismic data.

%\textbf{In this way, as an alternative to the classical scaling relations, we obtain new scaling relations for radius and mass by using MS and SG models constructed with {\small{MESA}} \citep{2011ApJS..192....3P,2013ApJS..208....4P,Paxton2015, Paxton2018, Paxton2019, Jermyn2023} evolutionary code.} 
{As an alternative to the classical scaling relations, we obtain new scaling relations based on reference frequencies for radius and mass by using MS and SG models constructed with {\small{MESA}} \citep{2011ApJS..192....3P,2013ApJS..208....4P,Paxton2015, Paxton2018, Paxton2019, Jermyn2023} evolutionary code.} 
In obtaining these relations, we derive simultaneous solutions for $\nu_{\rm min}$, $\braket{\Delta\nu}$, and $Z$ for specifically SG stars. We then apply the derived relations to \emph{Kepler} target stars.
Finally, we write a {\small{PYTHON}} code (version 3.13.2) \footnote{Python Software Foundation. Python Language Reference, version 3.13.2. Available at: \url{https://www.python.org}} to compute the mass and radius of all \emph{Kepler} target stars collectively.

{This paper is organized as follows.  In Section 2, we summarize the radius scaling relation from the literature and describe how the adopted seismic parameters are determined. In Section 3, we present the properties of the models constructed using the {\small MESA} and {\small CESAM} evolutionary codes. Section 4 is devoted to deriving new scaling relations for the mass and radius using the new methods. In Section 5, we apply the derived relations to \emph{Kepler} target stars and compare the results with values reported in the literature. Finally, the conclusions of our study are presented in Section 6. }
\section{Scaling relations in the literature}
Scaling relations are important for determining the fundamental parameters of stars by seismic methods.
%For this, we need some seismic parameters. In solar-like oscillating stars, the He II ionization zone causes small glitches at regular intervals and allows the formation of reference frequencies.
\citet{2019MNRAS.489.1753Y} used reference frequencies, in addition to the $\nu_{\rm max}$ and $\braket{\Delta\nu}$ parameters used in classical scaling relations. They computed the masses, radii, and ages of 90 stars by deriving new scaling relations, including relations based on $\nu_{\rm min}$. The radii of these stars computed from $\nu_{\rm min0}$ and $\nu_{\rm min1}$ are very close and the differences between them are less than 0.007 R$_{\odot}$.
%The scaling relations obtained for mass and radius are derived from MS models and applied to \emph{Kepler} target stars.
{The derived scaling relations were obtained from MS models and applied to \emph{Kepler} target stars.}
{However, these relations have certain dependencies and limitations. These scaling relations show a slight dependence on the metallicity.} Therefore, a new method was developed to compute the initial metallicity ($Z_{0}$) from the surface metallicity ($Z_{\rm s}$), taking into account the effect of microscopic diffusion. {The scaling relations are first derived based on $\braket{\Delta\nu}$ and $\nu_{\rm min}$, and then the effect of $Z$ is applied as a correction.}
The scaling relation for the radius based on $\nu_{\rm min1}$ ($R_{\rm sis1}$) from the MS models \citep{2019MNRAS.489.1753Y}:
\begin{equation}
{R_{\rm sis1}}/{R_{\odot}}=\frac{\left(\frac{\nu_{\rm min1}}{\nu_{\rm min1\odot}}\right)^{0.156} \left(\frac{\braket{\Delta\nu}_{\odot}}{\braket{\Delta\nu}}\right)^{0.92}}
{\left(1.14\left(r_{T\Gamma}-1.11\right)^{2}+0.98\right)\left(-0.64{r_{\delta\Delta}}+1.05\right)}.
    \label{eq:klasik_ölc_ilis_R}
\end{equation}
Here, the denominator includes terms involving the adiabatic exponent ($\Gamma_1$) and the small frequency separation ($\delta\nu_{02}$). $r_{T\Gamma}$ is defined as the ratio of $T_{\rm eff}$ to $\Gamma_1$ at the stellar surface ($r_{T \Gamma} = ({T_{\rm eff}}/{T_{\rm eff \odot}}) ({\Gamma_{1s \odot}}/{\Gamma_{1s}})$). The small separation ratio ($r_{\delta\Delta}$) is defined as the ratio of $\braket{\delta\nu_{02}}$ to $\braket{\Delta\nu}$ ($r_{\delta \Delta}={\braket{\delta\nu_{02}}}/{\braket{\Delta\nu}}$).
{A major limitation in applying this relation to certain model grids or observations is that the $\Gamma_1$ and $\delta\nu_{02}$ data cannot always be determined. When these terms are neglected, the relation reduces to a form that retains only the leading terms:}
\begin{equation}
\frac{R_{\rm sis1}}{R_{\odot}} {\approx} \left(\frac{\nu_{\rm min1}}{\nu_{\rm min1\odot}}\right)^{0.156} \left(\frac{\braket{\Delta\nu}_{\odot}}{\braket{\Delta\nu}}\right)^{0.92}.
    \label{eq:ölcek_iliksi_r_2019}
\end{equation}
%By expressing $\nu_{\rm min1}$ in function of $\nu_{\rm min0}$ ($\nu_{\rm min1}\propto\nu_{\rm min0}^{1.042}$), they obtain an equation similar to the one derived for $R_{\rm sis1}$ for $R_{\rm sis0}$.  
By expressing $\nu_{\rm min1}$ as a function of $\nu_{\rm min0}$ ($\nu_{\rm min1}\propto\nu_{\rm min0}^{1.042}$), they obtained a similar equation for $R_{\rm sis0}$.
%This section aims to explain why the scaling relations adopted here are formulated differently from those used in previous studies. 
%The new scaling relations derived in this study are based on equation (\ref{eq:ölcek_iliksi_r_2019}). Although this relation does not explicitly include metallicity, new relations incorporating metallicity effects have been derived by introducing the parameter $Z$.
{Using the $\nu_{\rm min}$ and $\braket{\Delta\nu}$ terms in equation (\ref{eq:ölcek_iliksi_r_2019}), we derive new scaling relations by extending the formulation to include additional parameters such as $Z$.}

For SG stars, several terms appearing in equation (\ref{eq:klasik_ölc_ilis_R}) are known to become less reliable as stellar evolution proceeds. The formulation adopted in this study, therefore, offers important advantages for evolved stars. It avoids an explicit dependence on the small frequency separation $\delta\nu_{02}$, which becomes increasingly difficult to determine with evolution, and includes the effect of metallicity directly in the scaling relation rather than applying it as a subsequent correction. This approach aims to provide more consistent and applicable results, particularly for stars in the SG evolutionary phase.

{Moreover, excluding quantities such as $\Gamma_{1s}$ from the scaling relations is not a limitation but may be a significant improvement for SG stars. The parameter $\Gamma_{1s}$ is typically derived from model-based fits as a function of $T_{\rm eff}$, which may not be well calibrated for the lower $T_{\rm eff}$ characteristic of SG stars and may depend on several modelling assumptions. By using only directly observable parameters, the scaling relations adopted in this study provide a robust framework for evolved stars.
The discussion in this section is therefore limited to outlining the physical motivation for the adopted scaling relations, while model tests and quantitative comparisons are presented in the relevant analysis sections.}
\subsection{Determination of seismic parameters} \label{sec2.1}
To obtain the stellar mass and radius using the scaling relations, we need to determine the seismic parameters. %$\Delta\nu$ and $\nu_{\rm {min}}$. 
$\Delta\nu$ represents the frequency difference between oscillation modes of the same angular degree (\emph{l}) but consecutive radial orders (\emph{n}): $\Delta\nu$ = $\nu_{n,l} - \nu_{n-1,l}$.
As shown in equation (\ref{eq:ölcek_iliksi_r_2019}), the other seismic parameter required to determine the stellar radius is $\nu_{\rm min}$.
{For this purpose, the reference frequencies are determined using $\Delta\nu$–$\nu$ diagrams constructed based on interior models.}
%using Gnuplot (e.g., version 5.4).} \footnote{Gnuplot version 5.4. Available at: \url{http://www.gnuplot.info/}}.
{To determine the frequency range of the model data, the observed frequencies of \emph{Kepler} target stars are used. The distribution of these observed frequencies with respect to $\nu_{\rm {max}}$ are examined. It was found that the frequency range of these stars is distributed approximately 22 per cent below and above $\nu_{\rm {max}}$. Considering this distribution, after calculating $\nu_{\rm {max}}$ values for the models, the lower and upper frequency limits are determined. Then, $\braket{\Delta\nu}$ and $\braket{\delta\nu_{02}}$ values of the models are calculated. {The $\braket{\delta\nu_{02}}$ is used to determine the evolutionary stage of the star and is not included in the scaling relations.}}
\begin{figure}
  \centering \includegraphics[width=\columnwidth]{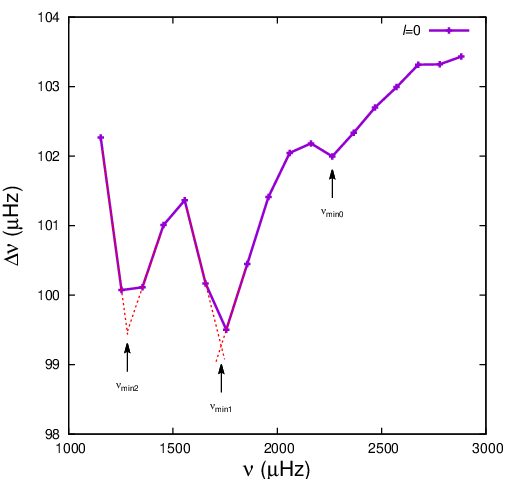}
  \caption{Variation of $\Delta\nu$ as a function of $\nu$. The degree of the modes is $l=0$. The linear equations describing the two slopes are shown with dashed points. The intersection of these dashed lines gives us $\nu_{\rm min1}$.} 
  \label{fig:ref_req_8645}
\end{figure}

%To determine the frequency corresponding to a specific minimum, it is sufficient to derive the linear equations {($f(x) = ax + b$) }that describe the slopes on either side of the dip.
The variation of $\Delta\nu$ with $\nu$ for the {\small{CESAM}} 1.0 M$_\odot$ SG model (first model in Table \ref{tab:CESAM_aks}) is plotted in Fig. \ref{fig:ref_req_8645} for $l = 0$.  The ranges of $\nu$ and $\Delta\nu$ are 1150–2900 $\mu$Hz and 103.4–99.5 $\mu$Hz, respectively. 
{The reference frequencies may in some cases correspond to a well-defined dip, as in the case of $\nu_{\rm min0}$ shown in Fig. \ref{fig:ref_req_8645}. When it cannot be directly determined (e.g. $\nu_{\rm min1}$ and $\nu_{\rm min2}$), the dip may appear flatter or less well-defined. In such cases, to determine the frequency corresponding to a specific minimum, it is sufficient to derive the linear equations {($f(x) = ax + b$)} that describe the slopes on either side of the dip. The point where these two lines intersect gives the frequency of $\nu_{\rm min}$.} 

In Fig. \ref{fig:ref_req_8645}, the red dashed lines represent these linear equations, and $\nu_{\rm min0}$ corresponds to 2263.97 $\mu$Hz, $\nu_{\rm min1}$ to 1731.75 $\mu$Hz, and $\nu_{\rm min2}$ to 1281.16 $\mu$Hz. In general, $\nu_{\rm min1}$ is much deeper than $\nu_{\rm min0}$. This allows a better determination of $\nu_{\rm min1}$ among the reference frequencies. Therefore, it is expected that the radii computed with $\nu_{\rm min1}$ will give more accurate results. We determine $\nu_{\rm min0}$, $\nu_{\rm min1}$, and $\nu_{\rm min2}$ from the oscillation frequencies of the models using this method.
The uncertainty in the reference frequencies, especially for $\nu_{\rm min1}$ and $\nu_{\rm min2}$, is very small compared to $\braket{\Delta\nu}$ in most cases, about one tenth of $\braket{\Delta\nu}$.

\begin{figure}
  \centering  \includegraphics[width=\columnwidth]{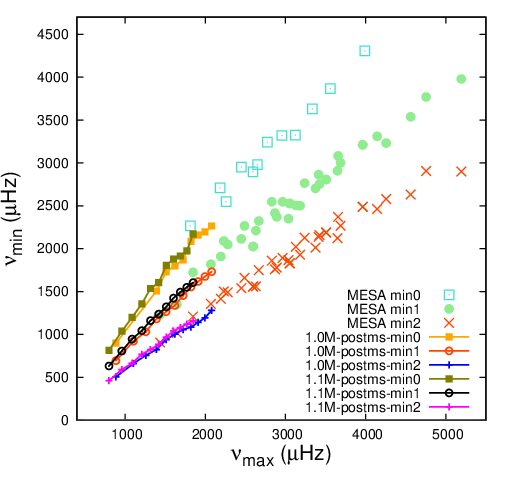}
  \caption{$\nu_{\rm max}$ versus $\nu_{\rm min}$ for 1.0 and 1.1 M$_\odot$ SG models ({\small CESAM}) with different masses computed in this study. The reference frequencies are compared with values obtained from {\small MESA} models taken from the study by \protect\cite{2016MNRAS.462.1577Y}. Squares, filled circles, and crosses represent $\nu_{\rm min0}$, $\nu_{\rm min1}$, and $\nu_{\rm min2}$ of the {\small MESA} model frequencies, respectively.} 
  \label{numax_krs_numin_pms}
\end{figure}

Most stars have many minima in both their observation and model oscillation frequencies. The numbering of these minima is very important. 
{The reference frequencies obtained from interior models are plotted versus $\nu_{\rm max}$ in Fig.~\ref{numax_krs_numin_pms}. To assess the consistency of the reference frequencies, the {\small{MESA}} $\nu_{\rm min}$ values taken from \citet{2016MNRAS.462.1577Y} are compared with the results obtained from the 1.0 and 1.1 M$_\odot$ SG interior models ({\small CESAM}) used in this study (see Tables \ref{tab:CESAM_ak} and \ref{tab:CESAM_aks}).}
%\textbf{The reference frequencies obtained for the 1.0 and 1.1 M$_{\odot}$ subgiant models indicate minima that are consistent with the results from the {\small MESA} models.}

{%Another seismic parameter that provides important information about the stellar interior, 
$\delta\nu_{02}$ is determined using the asymptotic relation \citep{1980ApJS...43..469T,1988Natur.336..634C}.} It represents the frequency difference between modes with angular degree $l=0$ and $l=2$ of consecutive radial orders: $\delta\nu_{02}(n)$ = $\nu_{n,0} - \nu_{n-1,2}$. Here, \emph{l} is related to the depth that the sound wave can reach.
When \emph{l}=0, the wave passes through the core, whereas for \emph{l}=2, the wave is reflected and returns from the core’s outer layers. In MS stars, the value of $\delta\nu_{02}$ is generally large, which means the frequency differences between \emph{l}=0 and \emph{l}=2 are noticeable. The $\delta\nu_{02}$ provides information about the evolutionary stage. 
{Specifically, it is directly related to the conversion of hydrogen into helium at the core.} 
%As hydrogen is converted into helium, the mean molecular weight ($\mu$) increases, which reduces the sound speed, since $c_{s} \propto \sqrt{P/ \rho}$ or $c_{s} \propto \sqrt{T/ \mu}$. As a result, the oscillation frequency differences decrease during MS evolution, leading to a decrease in $\delta\nu_{02}$ as the star evolves.}
%Thus, although $c_{s}$ increases with pressure ($P$) and temperature ($T$), it decreases with increasing density ($\rho$) and $\mu$.}
%Specifically, it is directly related to the conversion of hydrogen into helium at the core. When hydrogen in the core is converted to helium, the average molecular weight ($\mu$) increases. As $\mu$ increases, the speed of sound ($c_{s}$) of the star decreases, because the $c_{s}$ is given by: $c_{s} \propto \sqrt{P/ \rho}$ or $c_{s} \propto \sqrt{T/ \mu}$. The $c_{s}$ increases with pressure ($P$) and temperature ($T$) but decreases with increasing density ($\rho$) and $\mu$.

%As a result, the difference between the oscillation frequencies decreases throughout the MS evolution, and a decrease in  $\delta\nu_{02}$ is observed. This leads to a smaller frequency difference between \emph{l}=0 and \emph{l}=2 in evolved stars.

Furthermore, as the star evolves, its outer layers expand and the core becomes more compact. The oscillation properties associated with the interior structure change significantly, and mixed modes, particularly in higher degree modes such as $l=1$ and $l=2$, become dominant. 
As a result of these structural changes, the interpretation of $\delta\nu_{02}$ is more difficult compared to the MS stars.
%As a result of these structural changes, the effect of $\Gamma_{1}$ on the oscillation frequencies decreases, making the interpretation of $\delta\nu_{02}$ more difficult compared to MS stars. 
Therefore, the $R_{\rm sis1}$ relation (equation \ref{eq:ölcek_iliksi_r_2019}) for SG stars yields much more consistent results.
The $\delta\nu_{02}$ parameter continues to provide structural information for post-MS stars, as it does for MS stars. However, for the reasons explained above, its influence is weaker than in MS stars. Recent studies have derived new expressions for $\delta\nu_{02}$ that represent both MS and red giant stars \citep{2025ApJ...980..199O}. 
Consequently, $\delta\nu_{02}$ is not as effective for determining the $M$ and $R$ of evolved stars. Therefore, the relations derived in this study yield significantly better results for post-MS stars compared to MS stars.
\section{Models constructed by using the {\small{MESA}} and {\small{CESAM}} code}
%Interior models are derived by solving four non-linear stellar structure equations: the conservation of mass, hydrostatic equilibrium, thermal balance and temperature gradient.
{In this study, scaling relations for the stellar radius and mass are derived using MS and SG models constructed with {\small MESA}. The resulting relations are tested on {\small CESAM} models \citep[][see Appendix \ref{App_A}]{Marques2008}.}

%The {\small{MESA}} models are taken from \cite{2019MNRAS.490.1509K} and the {\small CESAM} models from \cite{2008Ap&SS.316..173M}.

%\textbf{The new scaling relations for the radius and mass from the models are constructed using the {\small{MESA}} evolution code.}
\subsection{ {\small{MESA}} models}
The {\small{MESA}} evolution code integrates many numerical and physics modules for a variety of stellar evolution scenarios, including advanced phases of star evolution from low-mass stars to massive stars. The models used in this study are taken from \cite{2019MNRAS.490.1509K}, which includes 20 host stars. Interior models were constructed using the observational data of 20 stars as constraints. Both asteroseismic and non-asteroseismic parameters were fitted by treating the initial helium abundance of the models ($Y_{\rm 0mod}$) and the initial mass of the models ($M_{\rm 0mod}$) as free parameters. These models cover the MS mass range of 0.74–1.27 M$_{\odot}$ and SG mass range of 1.07–1.55 M$_{\odot}$. The corresponding radius ranges for these masses are 0.75–1.42 R$_{\odot}$ and 1.25–2.10 R$_{\odot}$, respectively. $Z$ of these models ranges from 0.012 to 0.026. These models' $Z_0$ value ($Z_{\rm 0mod}$) was calculated using the formula $Z_{\rm 0mod} = 10^{[\rm M/H]_{obs}}Z_{0 \odot}$. 
%$Y_{\rm 0mod}$ was properly changed to fit the asteroseismic and non-asteroseismic constraints. 
{The models have $Y_{\rm 0mod}$ ranging between 0.261 and 0.292.}
Standard mixing length theory \citep{1958ZA.....46..108B} was assumed for convection theory in these models. For opacity at high temperatures, OPAL \citep{1993ApJ...412..752I,1996ApJ...464..943I} tables and low-temperature tables of \cite{2005ApJ...623..585F} were used. Nuclear reaction rates are computed using \cite{1999NuPhA.656....3A} updated by \cite{2002ApJ...567..643K} and \cite{2010ApJS..189..240C}. For atmospheric conditions in the host star models, the \texttt{simple\_photosphere} option is selected in {\small{MESA}}. Diffusion models have been developed for host stars with masses $M_{\rm star}$ < 1.2 M$_{\odot}$. Diffusion is not considered for stars larger than this mass.

The oscillation frequencies were derived using the ADIPLS package \citep{2008Ap&SS.316..113C} within {\small{MESA}}. Model $\nu_{\rm max}$ value was determined from the relation in \cite{1991ApJ...368..599B}. 
%Solar values ($\nu_{\rm max\odot}$ = 3050 $\mu$Hz \citep{1989sun..book.....S} and T$_{\rm eff\odot}$ = 5777 K \citep{1995A&A...293...87K}) were used in the calculations. 
$\Delta\nu$ and $\delta\nu_{02}$ were computed from the model oscillation frequencies. The $\nu_{\rm min}$ of the models were determined using the method described in \cite{2014MNRAS.441.2148Y}. 
\subsection{{\small{CESAM}} models}

{The derived scaling relations are tested using stellar evolution model grids \citep{2008Ap&SS.316..173M} constructed with the {\small CESAM} code \citep{1997A&AS..124..597M,2008Ap&SS.316...61M}, version 2k (see Appendix \ref{App_A}).} A new version of {\small{CESAM}} developed by \cite{Marques2008} was used to generate grids with different initial parameters. 
The zero-age main sequence (ZAMS) is defined as the point where 99 per cent of the star’s total energy is produced by nuclear reactions, while the terminal-age MS is defined as the point where the central hydrogen abundance $X_c= 0.01 \pm 0.0001$.
The models extend from the ZAMS to the beginning of the red giant branch. They are no-diffusion models with masses ranging from 0.8 to 8 M$_{\odot}$ and do not take into account the overshooting of the convective core. In the {\small{CESAM}} models, grids were constructed in the mass range 0.8--2.0 M$_{\odot}$ with steps of 0.1 M$_{\odot}$. From these grids, the 1.0 and 1.1 M$_{\odot}$ models with available seismic data are used in this study. The chemical composition (Y = 0.28 and Z/X = 0.02857) is the solar composition. In the calculations, an isothermal atmosphere was assumed at the top. OPAL tables \citep{1996ApJ...456..902R,1996ApJ...464..943I,1994ApJ...437..879A} were used for opacity at different temperatures. Models were computed with the GN93 solar mixture of heavy elements \citep[$(Z/X)_\odot$ = 0.0245;][]{1993oee..conf...15G}. Standard mixing length treatment of \cite{1958ZA.....46..108B} and \cite{1965ApJ...142..841H} is assumed for the convection theory.

In this study, we use the 1.0 and 1.1 M$_{\odot}$ {\small{CESAM}} models, for which the frequency files are available at www.astro.up.pt/helas/stars/cesam/. The Porto Oscillations Code (POSC) \citep{2008Ap&SS.316..121M} was used to compute the oscillation frequencies of these models. 
The frequency files (files of type {\tt.freq}) of the MS models contain oscillation frequencies for modes in the range 0 $\leq \emph{l} \leq$ 3. Although some post-MS models have frequencies computed for the same range, in general, the {\tt.freq} files contain only the \emph{l}=0 mode frequencies. The radius range covered by the masses of these models used in this study is 0.91–1.32 R$_{\odot}$ for MS models and 1.22–2.13 R$_{\odot}$ for SG models.

\section{New scaling relation for radius and mass with reference frequencies} \label{sec4} 
Determining the mass and radius is crucial for understanding the stellar interior structure and evolution. 
%{\small{CESAM}} model data is particularly useful for radius analysis (see Appendix \ref{App_A}). \textbf{The model data together with determined seismic parameters are listed in Tables \ref{tab:CESAM_ak} and \ref{tab:CESAM_aks}.} However, since the mass range is narrow, {\small{MESA}} models are used in mass \textbf{and radius} analysis. 
{The asteroseismic and model parameters determined using the oscillation frequencies are taken from \cite{2019MNRAS.490.1509K}. The models are divided into two groups: MS and SG.}
%The interior models obtained by \cite{2019MNRAS.490.1509K} for 20 host stars at different evolutionary phases using the {\small{MESA}} evolution code are very suitable for mass and radius analysis. These stars are solar-like oscillators and reference frequencies can be determined from observational data. The asteroseismic and model parameters determined using oscillation frequencies are taken from Tables 2 and 3 in their study. We divided the models into two different groups as MS and SG.

%In this section, we obtain scaling relations for R and M based on the reference frequencies from the MS and SG models.
{In the derivation of the scaling relations for R and M, a logarithmic method (equation \ref{eq:ms_loglogRsis}) is applied.
%In the study of \cite{2019MNRAS.489.1753Y}, the scaling relations for the mass and radius of the MS stars are determined. 
While deriving the fitting formulas for the $R$ and $g$, seismic parameters ($\nu_{\rm min}$ and $\braket{\Delta\nu}$) are analysed.
By extending the scaling relations from MS stars to SG stars, we simultaneously determine the exponents and coefficients of both $\nu_{\rm min}$ and $\langle \Delta\nu \rangle$.}

We can write equation (\ref{eq:ölcek_iliksi_r_2019}) in a more generalized scaling form to determine the radius relation:
\begin{equation}
\frac{R_{\rm sis}}{R_{\odot}}=10^{a}\left(\frac{\nu_{\rm min}}{\nu_{\rm min\odot}}\right)^{b} \left(\frac{\Delta\nu_{\odot}}{\Delta\nu}\right)^{c}.
    \label{eq:ms_log_Rsis10}
\end{equation}
The logarithmic form of this equation is as follows:
\begin{equation}
\log(\frac{R_{\rm sis}}{R_{\odot}})=a+b\log\left(\frac{\nu_{\rm min}}{\nu_{\rm min\odot}}\right)+c\log \left(\frac{\Delta\nu_{\odot}}{\Delta\nu}\right).
    \label{eq:ms_loglogRsis}
\end{equation}
Equation (\ref{eq:ms_loglogRsis}) is an equation with two variables and three unknowns:
\begin{equation}
f(x,y)=a+bx+cy
    \label{eq:f(x,y)_fonk}.
\end{equation} 
In the derived relations, the solar seismic values listed in Table \ref{sismik_gunes_parametreleri} are used.
\begin{table}
  \centering
  \caption{Asteroseismic solar parameters. $\nu_{\rm max}$ and $\Delta\nu_\odot$ are from \citet{huber.2011}, the reference frequencies and $\delta\nu_{{02}\odot}$ is from \citet{2016MNRAS.462.1577Y}.}
    \begin{tabular}{lllllllll}
    \hline
    \multicolumn{3}{l}{Solar asteroseismic parameters} & \multicolumn{3}{c}{} & \multicolumn{3}{l}{References} \\
    \hline
    \multicolumn{3}{l}{$\nu_{\rm max}$$_\odot$} & \multicolumn{3}{c}{3090 $\pm$ 30 $\mu$Hz} & \multicolumn{3}{l}{(Huber et al. 2011)} \\
    \multicolumn{3}{l}{$\braket{\Delta\nu_\odot}$} & \multicolumn{3}{c}{135.1 $\pm$ 0.1 $\mu$Hz} & \multicolumn{3}{l}{} \\
    \multicolumn{3}{l}
    {$\nu_{\rm min0}$} & \multicolumn{3}{c}{3256.6 $\pm$ 32.6 $\mu$Hz} & \multicolumn{3}{l}{(Yıldız et al. 2016)} \\
    \multicolumn{3}{l}
    {$\nu_{\rm min1}$} & \multicolumn{3}{c}{2555.2 $\pm$ 25.6 $\mu$Hz} & \multicolumn{3}{l}{} \\
    \multicolumn{3}{l}
    {$\nu_{\rm min2}$} & \multicolumn{3}{c}{1879.5 $\pm$  18.8 $\mu$Hz} & \multicolumn{3}{l}{} \\
    \multicolumn{3}{l}{$\braket{\delta\nu_{02}}$$_\odot$} & \multicolumn{3}{c}{9.8 $\pm$ 0 $\mu$Hz} & \multicolumn{3}{l}{} \\
    \hline
    \end{tabular}%
  \label{sismik_gunes_parametreleri}%
\end{table}%

Highly accurate, direct relations based on reference frequencies cannot be obtained for the mass. In order to obtain an explanation for the mass, first of all, it is necessary to determine \emph{g} and \emph{R} well. We can compute the seismic mass ($M_{\rm sis}$) by deriving the relations for \emph{g} and \emph{R} from the seismic data of the models:
\begin{equation}
\frac{M_{\rm sis}}{M_{\odot}}=\left(\frac{g_{\rm sis}}{g_{\odot}}\right) \left(\frac{R_{\rm sis}}{R_{\odot}}\right)^{2}.
    \label{eq:ms_Msis1_2019}
\end{equation}
To determine the mass, we use {\small{MESA}} models in the mass range of 1.07-1.55 M$_{\odot}$. These models include SG models. 
%Microscopic diffusion is included in the construction of models with $M<1.2$ $\rm M_{\odot}$. For $M>1.2$ $\rm M_{\odot}$, it is not included. 

The $Z_{\rm 0mod}$ parameter in table 2 of \cite{2019MNRAS.490.1509K} is $Z_0$ value.
The {\small{MESA}} models are constructed for a wider mass range and for different compositions.  
Model parameters such as $g$ and $R$ are also known to show a slight dependence on $Z$. Therefore, we use the three-variable function $f(x,y,t)$ instead of the two-variable $f(x,y)$ in equation~ (\ref{eq:f(x,y)_fonk}). For this, we define the parameter $t$:
\begin{equation}
t=\frac{Z_0}{Z_{\odot}}+0.1.
\end{equation}
The solar metallicity is taken to be $Z_{\odot} = 0.0134$ \citep{2009ARA&A..47..481A}. 
%We added the term 0.1 to account for stars with $Z$ values of 0 or very small. 
The term 0.1 is added to account for stars with a $Z$ value of 0 or very small values. Then, we separately derived the simultaneous solutions of three-variable ($\nu_{\rm min}$, $\braket{\Delta\nu}$ and $Z_{\rm mod}$) from the MS and SG models with {the logarithmic method} for ${g_{\rm log}}$ and ${R_{\rm log}}$:
\begin{equation}
\frac{g_{\rm log}}{g_{\odot}}=10^{k'}\left(\frac{\nu_{\rm min}}{\nu_{\rm min\odot}}\right)^{l'}\left(\frac{\Delta\nu_{\odot}}{\Delta\nu}\right)^{m'}\left(\frac{Z_0}{Z_{\odot}}+0.1\right)^{n'},
\label{eq:ms_Msis_g}
\end{equation}

\begin{equation}
\frac{R_{\rm log}}{R_{\odot}}=10^{k}\left(\frac{\nu_{\rm min}}{\nu_{\rm min\odot}}\right)^{l}\left(\frac{\Delta\nu_{\odot}}{\Delta\nu}\right)^{m}\left(\frac{Z_0}{Z_{\odot}}+0.1\right)^{n}.
\label{eq:ms_Msis_R}
\end{equation}

\begin{figure}
  \centering
\includegraphics[width=1.4\linewidth]{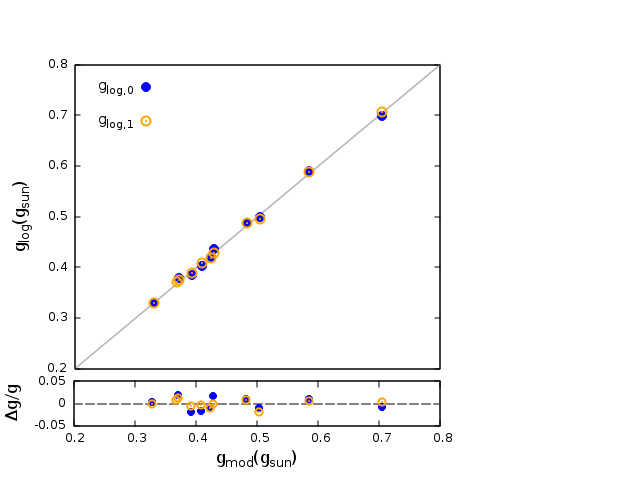}
  \caption{$g_{\rm log,0}$ and $g_{\rm log,1}$ are plotted with respect to $g_{\rm mod}$ for SG models. The filled circles are for the surface gravity computed for $\nu_{\rm min0}$ and the circles are for the surface gravity computed for $\nu_{\rm min1}$. {The bottom panel shows the differences in surface gravity, ranging from -0.02 to 0.02.}}
  \label{fig:gsis01.krs.gmod_POSTMS}
\end{figure}

\begin{figure}
  \centering    \includegraphics[width=1.4\linewidth]{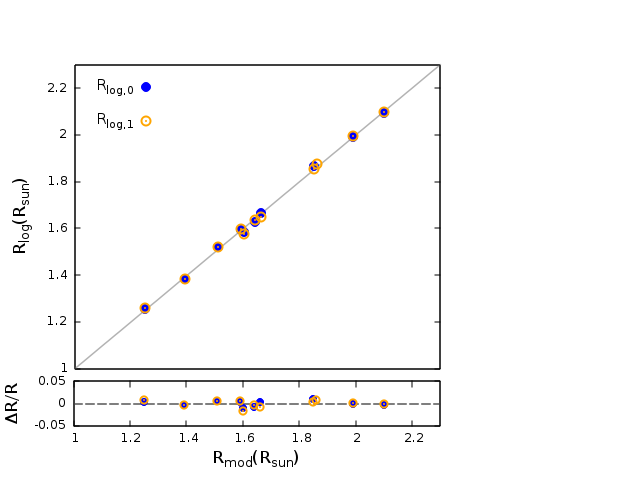}
  \caption{$R_{\rm log,0}$ and $R_{\rm log,1}$ are plotted with respect to $R_{\rm mod}$ for SG models.
  Filled circles indicate the radii computed for $\nu_{\rm min0}$, while circles indicate the radii computed for $\nu_{\rm min1}$. Radii range from 1.2-2.2 R$_{\odot}$. {The bottom panel shows the differences in radius, ranging from -0.016 to 0.010.}} 
  \label{fig:Rsis01.krs.Rmod_POSTMS}
\end{figure}

\begin{table}
  \centering
  \caption{Coefficients and uncertainties for $R$ and $g$ from {\small MESA} SG models, determined by logarithmic method.}
    \begin{tabular}{lrrrr}
    \hline
    Coefficients & \multicolumn{1}{c}{$R_{\rm log,0}$} & \multicolumn{1}{c}{$R_{\rm log,1}$} \\
    \hline
    $k$     & -0.0066 $\pm$ 0.0079 & -0.0071 $\pm$ 0.0100 \\
    $l$     & 0.191 $\pm$ 0.042 & 0.211 $\pm$ 0.062 \\
    $m$     & 0.964 $\pm$ 0.035 & 0.997 $\pm$ 0.054 \\
    $n$     & 0.063 $\pm$ 0.018 & 0.068 $\pm$ 0.020 \\
    \hline
          & \multicolumn{1}{c}{$g_{\rm log,0}$} & \multicolumn{1}{c}{$g_{\rm log,1}$} \\
    \hline
    $k'$    & -0.0004 $\pm$ 0.0179 & 0.0142 $\pm$ 0.0119 \\
    $l'$    & 0.346 $\pm$ 0.095 & 0.498 $\pm$ 0.074 \\
    $m'$    & -0.893 $\pm$ 0.078 & -0.731 $\pm$ 0.064 \\
    $n'$    & 0.039 $\pm$ 0.040  & 0.016 $\pm$ 0.024 \\
    \hline
    \end{tabular}%
  \label{tab:pms_Rlog_glog_coef}%
\end{table}%
{We used the {\small{MESA}} models introduced in the previous section to determine the best-fitting coefficients of the adopted scaling relations.} The coefficients for $R_{\rm log}$ and $g_{\rm log}$ along with their uncertainties, are listed in Table \ref{tab:pms_Rlog_glog_coef}. {The values listed in Table \ref{tab:pms_Rlog_glog_coef} correspond specifically to SG stars.} The model surface gravity ($g_{\rm mod}$) is plotted against $g_{\rm log}$ calculated from $\nu_{\rm min0}$ and $\nu_{\rm min1}$ in Fig. \ref{fig:gsis01.krs.gmod_POSTMS} and $R_{\rm mod}$ plotted against $R_{\rm log}$ in Fig. \ref{fig:Rsis01.krs.Rmod_POSTMS} are shown. The agreement of surface gravity and radii with the model values is very clear. This is crucial for the precise determination of the mass.
%The logarithmically derived relations for $g$ and $R$ show a good agreement with the MS models. However, the results obtained using $\nu_{\rm min1}$ provide a slightly better fit for both parameters. 
%Additionally, for the scaling relations of mass and radius in MS stars, see \cite{2019MNRAS.489.1753Y}.

\begin{figure}
  \centering        \includegraphics[width=1.4\linewidth]{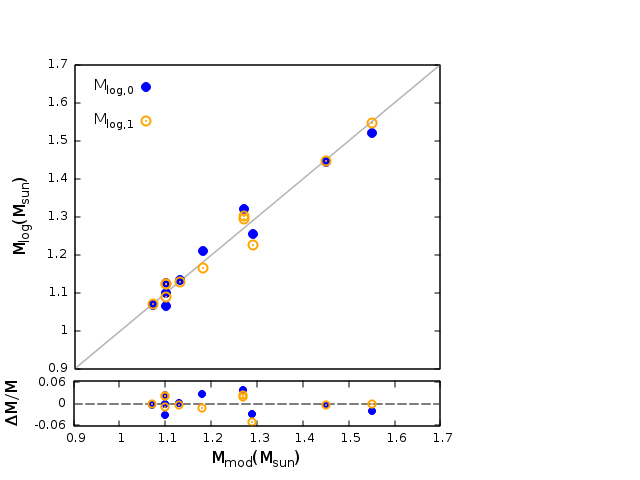}
  \caption{$M_{\rm log,0}$ and $M_{\rm log,1}$ are plotted with respect to $M_{\rm mod}$ for SG models. The filled circles represent the masses computed for $\nu_{\rm min0}$, while the open circles correspond to masses computed for $\nu_{\rm min1}$.
  The masses are
  plotted in the range of 1.01-1.55 M$_{\odot}$  in all {\small{MESA}} SG models. {The bottom panel shows the differences in $M$, ranging from -0.05 to 0.04.}}
  \label{fig:Msis01.krs.Mmod_POSTMS}
\end{figure}

%In this study, we mainly focus on the \emph{M} and \emph{R} relations of SG stars. 
{To test how well these relations recover the known stellar masses, the results are presented in Fig. \ref{fig:Msis01.krs.Mmod_POSTMS}.}
Based on this analysis, we present the first model-based derivation of the mass relation for SG stars. We compute the masses $M_{\rm log,0}$ and $M_{\rm log,1}$ using $\nu_{\rm min0}$ and $\nu_{\rm min1}$. The comparison of the model mass ($M_{\rm mod}$) between $M_{\rm log,0}$ and $M_{\rm log,1}$ is shown in Fig.~\ref{fig:Msis01.krs.Mmod_POSTMS}.
The masses and radii derived closely match the model values of SG stars, although some scatter is observed in Fig.~\ref{fig:Msis01.krs.Mmod_POSTMS}. {This scatter arises from the contribution of the uncertainties in $g$ ($e_g$) and $R$ ($e_R$) to the mass uncertainty ($e_M$): $eM/M = eg/g + 2eR/R$.}
%Rnin x2 kat etkisi var. Ama esas hataya katkı dg den geliyor çünkü grağiklere baktığımızda range i büyük 0.3 - 0.8 (0.3*2.5kat rangein ucu). Ama Rnin range aralığı daha dar 1.2 -2.2 (x2 kat=2.4 bile yok).
Although the relations obtained from $\nu_{\rm min0}$ and $\nu_{\rm min1}$ are quite similar, the results derived from $\nu_{\rm min1}$ provide a better improvement. Because in general, the amplitude around $\nu_{\rm min1}$ is much deeper than that around $\nu_{\rm min0}$, allowing for a more precise determination of $\nu_{\rm min1}$ between the reference frequencies. Therefore, we predict that the radii and masses calculated with $\nu_{\rm min1}$ will give more accurate results.
%The results derived from $g_{\rm log,1}$ and $M_{\rm log,1}$ are more consistent compared to those obtained from $\nu_{\rm min0}$.

{In addition, the scaling relations derived for mass and radius can be tested by comparing the scaling relation estimates with the corresponding model values, as shown in Figs \ref{fig:Rsis01.krs.Rmod_POSTMS} and \ref{fig:Msis01.krs.Mmod_POSTMS}. The radii derived from both minima are in good agreement, with differences in most of the model radii being less than 1 per cent. For masses, the differences are below 2.5 per cent for $M_{\rm log,1}$, while they remain below 3 per cent for $M_{\rm log,0}$. These values are significantly smaller than the typical systematic uncertainties of order $\cong 5$ per cent in radius and $\cong 10$ per cent in mass obtained for evolved stars using classical scaling relations { \citep[see][]{2013ARA&A..51..353C,2017ApJ...844..102H}.}}

{The higher precision of the new scaling relations compared to classical scaling relations mainly arises from the exclusion of the $\nu_{\max}$ parameter, which typically has larger uncertainties than other seismic parameters. Instead, our relations are based on well-defined reference frequencies, allowing stellar masses and radii to be determined with improved precision.}

{To test the model dependence of the {\small MESA}-calibrated scaling relations (equations \ref{eq:ms_Msis_g} and \ref{eq:ms_Msis_R}), {\small CESAM} models are used (see Appendix \ref{App_A}). The oscillation frequencies of the {\small{CESAM}} 1.0 and 1.1 M$_{\odot}$ models are analysed to determine $\braket{\Delta\nu}$, $\braket{\delta\nu{02}}$, and the reference frequencies. The calibrated scaling relations are then applied to the {\small CESAM} models, and the resulting radii and masses are directly compared with the corresponding model values. 
The results indicate that the radius estimates do not show a strong dependence on the adopted stellar evolution code, whereas the mass estimates exhibit differences of up to $\sim$ 7 per cent between different modelling codes, which may arise from model-dependent effects.}

%The results indicate that, within the explored mass range, the scaling relations do not exhibit a significant dependence on the adopted stellar evolution code.

%These comparisons show negligible systematic offsets, with mean fractional differences consistent with zero for both radius and mass. The scatter is $\cong 0.6-0.7\%$ for radius and $\cong 2\%$ for mass. These values are significantly smaller than the typical systematic uncertainties of order $\cong 5\%$ in radius and $\cong 10\%$ in mass obtained for evolved stars using classical scaling relations \citep[see][]{2013ARA&A..51..353C}.}

\section{Results and Discussions}
{In this section, we apply the scaling relations derived in the previous section to \emph{Kepler} targets. Using these relations, we can determine the values of $M$ and $R$ from observational parameters alone, without the need for detailed stellar models. In this context, although we compare the derived masses and radii with results available in the literature, our primary focus is on analysing the general trends and systematic behaviour of the classical, $\nu_{\rm min0}$-based, and $\nu_{\rm min1}$-based scaling relations across different evolutionary phases. Furthermore, we analyse the accuracy, consistency, and potential systematic deviations of these relations to identify which approach provides more reliable results within specific parameter ranges.}

{To provide a more comprehensive assessment, we adopt an approach that integrates stellar modelling with theoretical and observational constraints.} This allows us to compare the observed frequencies of the star with the frequencies obtained from models using different input parameters. %Furthermore, we can evaluate how well the $M$ and $R$ relations derived from models agree with the observed $M$ and $R$ values. 
Furthermore, we can assess the agreement between the $M$ and $R$ values calculated from model oscillation frequencies and those obtained from observational oscillation frequencies.
A previous study by \cite{2019MNRAS.489.1753Y} applied solutions derived from MS models to target stars. 
%In their method, scaling relations based on $\braket{\Delta\nu}$ and $\nu_{\rm min}$ are first derived, then a $Z$-correction is applied.
%In contrast, in this study, we obtain a simultaneous solution for the parameters $\braket{\Delta\nu}$, $\nu_{\rm min}$ and $Z_{0}$. 
%\textbf{In contrast, in this study, we obtain a simultaneous solution for the parameters $\braket{\Delta\nu}$, $\nu_{\rm min}$, and $Z_{0}$ without applying the $Z$-correction afterwards. We then apply the scaling relations, derived for the first time for SG stars from reference frequencies, to the \emph{Kepler} target stars.} 

The data for \emph{Kepler} target stars are taken from table A1 in \cite{2016MNRAS.462.1577Y} and table B1 in \cite{2019MNRAS.489.1753Y}, and the corresponding parameters are obtained from observational frequencies. In addition, the $R_{\rm sis}$ and $M_{\rm sis}$ values in table B1 were determined using scaling relations derived from $\nu_{\rm min}$.
The radii $R_{\rm sis0}$ and $R_{\rm sis1}$ in Table B1 are referred to as $R_{0}$ and $R_{1}$ in this study, respectively. Similarly, $M_{\rm sis0}$ is defined as $M_{0}$ and $M_{\rm sis1}$ as $M_{1}$. {%The evolutionary phase of the stars was determined based on their $\delta\nu_{{02}}$ values. %The small frequency separation $\delta\nu_{{02}}$ is sensitive to structural changes in the stellar core and therefore to stellar evolution. As the star evolves, $\delta\nu_{{02}}$ decreases due to changes in the sound speed profile and therefore provides complementary information on the evolutionary stage of MS and SG stars. 
%As noted in Section \ref{sec2}, changes in the internal structure affect the small frequency separation, which in turn helps distinguish the evolutionary stage of stars. 
The evolutionary classification of Kepler target stars was based on parameters such as $M$, $R$, and the $\delta\nu_{{02}}$. Stars with small frequency separations in the range of 5.7–10.6 $\mu$Hz were classified as MS, while those with 5.5 $\mu$Hz were classified as SG stars. While the scaling relations have the same functional form for MS and SG stars, their coefficients differ and are determined separately from evolutionary models appropriate to each phase. We discuss the results for each evolutionary phase separately.}

{The $Z_0$ values listed in table B1 of \cite{2019MNRAS.489.1753Y} were determined in that study as follows. For MS stars, $Z_0$ was obtained from the observed surface metallicity ($Z_s$) by applying a correction for microscopic diffusion as a function of stellar mass and age. For post-MS stars, the effect of the deepening convective envelope was considered and $Z_0$ was determined using a parameter defined in terms of $T_{\rm eff}$ and $g$.}

%In the previous section, we derived the mass and radius relations for SG stars for the first time using the logarithmic method. 
{The new masses and radii of the SG \emph{Kepler} targets are presented in Table \ref{tab:seismic_parameters_postMS_Kls} along with their uncertainties.}  
The uncertainties in mass and radius are estimated by Monte Carlo simulations using the observed values of $\braket{\Delta\nu}$, $\nu_{\rm min0}$, $\nu_{\rm min1}$, and $Z$. In the simulation, 10 000 synthetic data points are generated for each parameter within its observational uncertainty range, assuming a Gaussian distribution. For each star, 10 000 mass and radius values are computed based on these synthetic samples. The 1 $\sigma$ uncertainties in mass ($\sigma_M$) and radius ($\sigma_R$) are then derived as the standard deviations of these distributions.
\subsection{Radius of the \emph{Kepler} targets}
{In this section, we apply the new radius relation to the MS and SG target stars and compute $R_{\rm log,0}$ and $R_{\rm log,1}$. 
%Radii are obtained as $R_{\rm log,0}$ and $R_{\rm log,1}$. 
%In addition, we present, for the first time, the derivation of a $\nu_{\rm min}$-based radius relation specifically for SG stars.
}
\subsubsection{MS \emph{Kepler} targets}
\begin{figure}
  \centering    \includegraphics[width=1.4\linewidth]{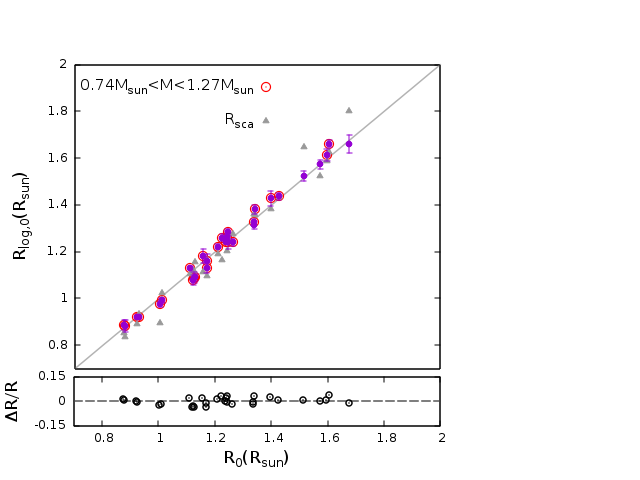}  \includegraphics[width=1.4\linewidth]{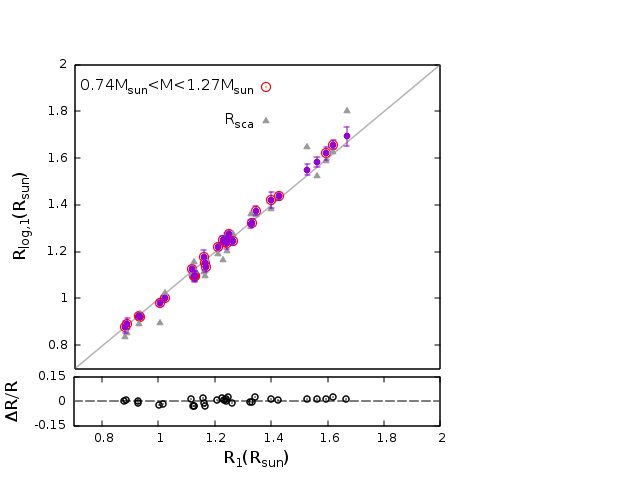}    \caption{$R_{\rm log}$ is plotted with respect to MS \emph{Kepler} target stars' radii. (a) $R_{\rm log,0}$ versus $R_{0}$. (b) $R_{\rm log,1}$ versus $R_{1}$. $R_{0}$ and $R_{1}$ are taken from \protect\cite{2019MNRAS.489.1753Y}. The red circles show target stars in the narrow mass range of 0.74-1.27 M$_{\odot}$ for which {\small{MESA}} models are available. The grey triangles indicate the radii determined using the classical scaling relation. {The bottom panels show the radius differences, which range from 0.04 to -0.04.}}
  \label{Kepler_ms_mesa_R01}
\end{figure}

%First, we apply the scaling relations for radius derived from the {\small{MESA}} MS models to \emph{Kepler} target stars. The relation between $R_{\rm log,0}$, determined using $\nu_{\rm min0}$, and $R_{0}$ is shown in Fig.~\ref{Kepler_ms_mesa_R01}a. Similarly, the relation between $R_{\rm log,1}$, obtained from $\nu_{\rm min1}$, and $R_{1}$ is presented in Fig.~\ref{Kepler_ms_mesa_R01}b. The triangles in Fig.\ref{Kepler_ms_mesa_R01} represent the radii determined from the classical scaling relation (equation \ref{eq:sca_rel_R}). 
The relation between $R_{\rm log,0}$, determined using $\nu_{\rm min0}$, and $R_{0}$ is shown in Fig.~\ref{Kepler_ms_mesa_R01}(a).{ Similarly, the relation between $R_{\rm log,1}$, obtained from $\nu_{\rm min1}$, and $R_{1}$ is presented in Fig.~\ref{Kepler_ms_mesa_R01}(b), along with a residual panel ($\Delta R/R$)}. The triangles in Fig.~\ref{Kepler_ms_mesa_R01} represent the radii determined from the classical scaling relation (equation \ref{eq:sca_rel_R}).  
{The radii obtained from $\nu_{\rm min0}$ and $\nu_{\rm min1}$ show consistent results.} 

{For MS stars, the radii obtained from equation (\ref{eq:ms_Msis_R}) show differences in their distribution compared to $R_{\rm sca}$. $R_{\rm log}$ exhibits a lower scatter, particularly in the ranges of 0.85–1.0~R$_\odot$ and 1.5–1.7~R$_\odot$, where the scatter in $R_{\rm sca}$ is large.}

%For MS stars, the radii obtained from equation (\ref{eq:ms_Msis_R}) show an overall improvement compared to $R_{\rm sca}$. $R_{\rm log}$ yields more consistent results, particularly in the ranges of 0.85–1.0~R$_\odot$ and 1.5–1.7~R$_\odot$, where the scatter in $R_{\rm sca}$ is large.

%This improvement is particularly evident in the ranges of 0.85–1.0~R$_\odot$ and 1.5–1.7~R$_\odot$, where the scatter in $R_{\rm sca}$ is large.
%\textbf{To quantitatively evaluate the agreement between the radii determined from the $\nu_{\rm min0}$ and $\nu_{\rm min1}$, we calculate the Pearson correlation coefficient ($r$), together with the mean and RMS of the $\Delta R / R$ values.For the MS stars, the results for radii derived from $\nu_{\rm min0}$ are $r = 0.9939$, $\langle \Delta R /R\rangle = 0.0008$ and $\mathrm{RMS} = 0.0208$. Correspondingly, for radii derived from $\nu_{\rm min1}$, we find $r = 0.9972$, $\langle \Delta R/R \rangle = 0.0018$ and $\mathrm{RMS} = 0.0165$.}Both radii determined from the $\nu_{\rm min0}$ and $\nu_{\rm min1}$ show a strong correlation with $R_{0}$ and $R_{1}$, and the near-zero mean residuals indicate no significant systematic bias. The radii derived from $\nu_{\rm min1}$ show a slightly better agreement, with a higher correlation coefficient and a lower RMS difference. A residual panel ($\Delta R/R$) showing these differences has been added to Fig. \ref{Kepler_ms_mesa_R01}.

\subsubsection{SG \emph{Kepler} targets}
\begin{figure}
  \centering    \includegraphics[width=1.4\linewidth]{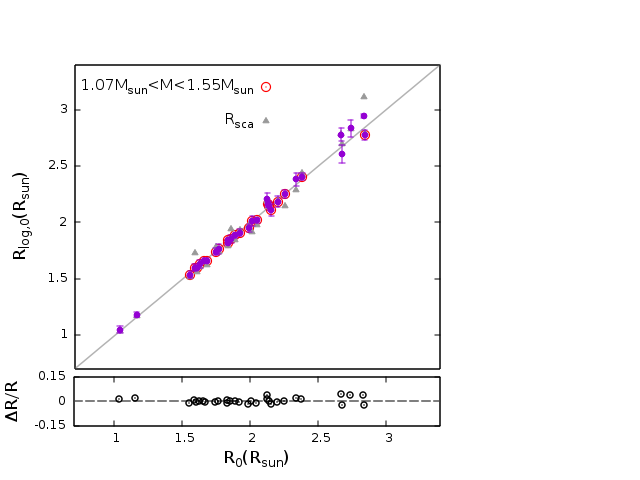}     \includegraphics[width=1.4\linewidth]{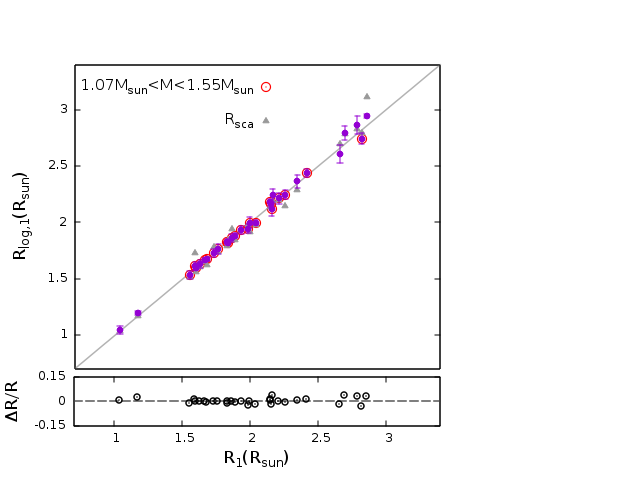}
  \caption{R$_{\rm log}$ versus SG \emph{Kepler} targets stars radii. Radii are plotted logarithmically. (a) R$_{\rm log,0}$ is plotted with respect to $R_{0}$. (b) $R_{\rm log,1}$ is plotted with respect to $R_{1}$. $R_{0}$ and $R_{1}$ are taken from \protect\cite{2019MNRAS.489.1753Y}. The circles show target stars in the mass range of $1.07$--$1.55$ M$_{\odot}$ for which {\small MESA} models are available. The triangles indicate the radii determined using the classical scaling relation. {The bottom panels show the radius differences, which range from 0.04 to -0.04.}}
  \label{mesa_postms_KepLegs}
\end{figure}

%We apply the radius relations for $\nu_{\rm min0}$ and $\nu_{\rm min1}$, derived from SG models, to \emph{Kepler} target stars. 
The radii determined from $\nu_{\rm min0}$ and $\nu_{\rm min1}$ are listed in Table \ref{tab:seismic_parameters_postMS_Kls}. The resulting radii, $R_{\rm log,0}$ and $R_{\rm log,1}$, are calculated from equation (\ref{eq:ms_Msis_R}). The determined $R_{\rm log}$ radii range from 1.04 to 2.95 R$_\odot$.
The scaling relations for $R$, derived from {\small{MESA}} SG models covering a wide mass range, are presented in Section \ref{sec4}. Based on these relations, the SG radii $R_0$ and $R_1$, corresponding to the computed $R_{\rm log,0}$ and $R_{\rm log,1}$ for $\nu_{\rm min0}$ and $\nu_{\rm min1}$, are plotted in Fig. \ref{mesa_postms_KepLegs}, respectively. A residual panel showing $\Delta R / R$ is also plotted in Fig. \ref{mesa_postms_KepLegs}. {The obtained radii are broadly consistent with the target stars' radii.}
%The obtained radii appear to be in good agreement with the target stars' radii. 
%Furthermore, for SG stars, the radii derived from $R_{\rm sca}$ and $R_{\rm log}$ are generally similar; however, the radii obtained from $R_{\rm log}$ provide a noticeable improvement. The agreement is particularly good in the range of 1–2.4~R$_\odot$, while some scatter is observed beyond 2.4~R$_\odot$.

%To further quantify the performance of the different scaling relations in the investigated mass and radius range, we analyse the residuals, $\Delta R/R$.
{For SG stars, we find that the results obtained from $\nu_{\rm min0}$ and $\nu_{\rm min1}$ are very similar, indicating that both minima yield consistent measurements. {Although the radii derived from $R_{\rm sca}$ and $R_{\rm log}$ are generally comparable, those from $R_{\rm log}$ show some differences in their distribution.}}

{One of the factors that may affect the scaling relations is $Z$. Therefore, the effect of $Z$ on the scaling relations is investigated (see Appendix \ref{App_B}). For this purpose, an alternative scaling relation for the radius is derived. This relation does not include a $Z$ term and is based on {\small MESA} SG models. The resulting radii are then compared. The analysis shows that $Z$ provides a very small, yet noticeable, improvement in the scaling relations.}
%\textbf{For SG stars, we find that the dispersion of the relations based on $\nu_{\rm min0}$ is larger than that of those based on $\nu_{\rm min1}$. This trend indicates that the deviations are not random but instead correlate with the evolutionary phase.As a star evolves, the expansion of its envelope modifies the internal density structure and sound-speed profile. These structural changes shift the frequencies corresponding to the minima toward lower values and modify the structure of the $\Delta\nu$-$\nu$ pattern \citep[e.g. see][]{2014MNRAS.441.2148Y}. In SG stars, the $\nu_{\rm min1}$ is more clearly and deeply defined than $\nu_{\rm min0}$, leading to more stable measurements and, consequently, more reliable results.}
\subsection{Mass of the \emph{Kepler} targets}
In this section, we apply the $M_{\rm log}$ solutions, derived from {\small{MESA}} MS and SG models, to the target stars. The description for mass determination from {\small{MESA}} models is given in Section \ref{sec4}. The results obtained from the MS models are compared with previous studies.
%As a key contribution of this study, we derive a scaling relation for mass based on SG models.The results obtained from the MS models are compared with previous studies.
\subsubsection{MS \emph{Kepler} targets}
\begin{figure}
  \centering   \includegraphics[width=1.4\linewidth]{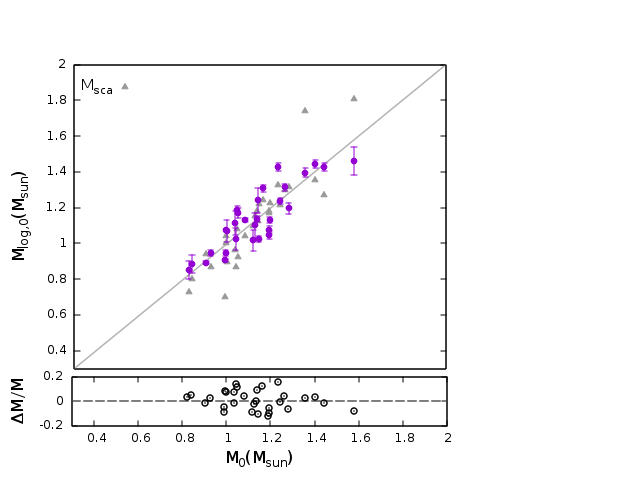}  \includegraphics[width=1.4\linewidth]{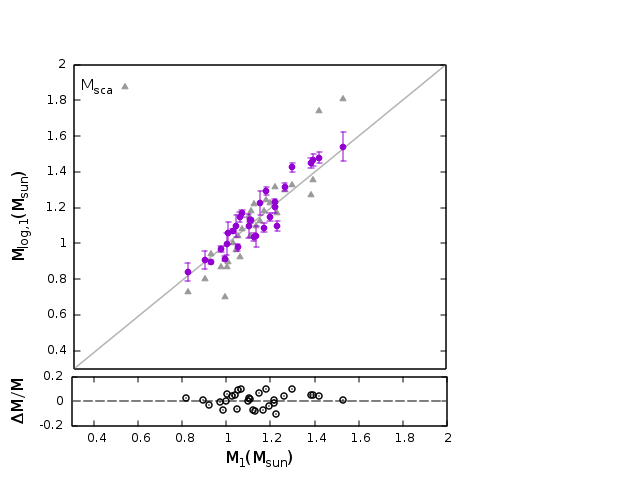}
  \caption{$M_{\rm log}$ versus \emph{Kepler} target stars masses for MS stars. (a) $M_{\rm log,0}$ is plotted with respect to $M_{0}$. (b) $M_{\rm log,1}$ is plotted with respect to $M_{1}$.  $M_{0}$ and $M_{1}$ are taken from \protect\cite{2019MNRAS.489.1753Y}. The triangles indicate the masses determined using the classical scaling relation. {The bottom panels represent the differences in mass, corresponding to $\pm 0.15$ in panel (a) and $\pm 0.10$ in panel (b).}}
  \label{ms_Mlog01_M01}
\end{figure}

The $M_{\rm log}$ solutions are applied to the MS target stars to compute the corresponding masses, $M_{\rm log,0}$ and $M_{\rm log,1}$. 
The relations between these model-based masses and the target stars' masses are as follows:
 \begin{equation}
M_{\rm log,0}=(1.0060\pm 0.0014){M_{0}},
\label{eq:Mlog0_M0_ms} 
\end{equation} 
 \begin{equation}
M_{\rm log,1}=(1.0140\pm 0.0011){M_{1}}.
\label{eq:Mlog1_M1_ms} 
\end{equation} 
The results show high consistency for both solutions. Since the multiplier of $M_{0}$ (1.006) is very close to unity, $M_{\rm log,0}$ shows significantly better agreement with $M_{0}$. However, when uncertainties are taken into account, $M_{\rm log,1}$ appears to yield more reliable results. Accordingly, we can predict that the relations for the radii will be much better. $M_{\rm log,0}$ versus $M_{0}$ is plotted in Fig.~\ref{ms_Mlog01_M01}(a) and $M_{\rm log,1}$ versus $M_{1}$ is plotted in Fig.~\ref{ms_Mlog01_M01}(b). The residual panel $\Delta M / M$ against the \emph{Kepler} mass is also plotted in Fig.~\ref{ms_Mlog01_M01}. The triangles represent the masses determined from the classical scaling relation. 
%(see equation~\ref{eq:sca_rel_R}).

$M_{\rm sca}$ and $M_{\rm log}$ yield very similar results within the 1.05–1.3 M$_{\odot}$ mass range. {In this intermediate region, the difference $M_{\rm log1} - M_{1}$ reaches its maximum as shown in Fig.~\ref{ms_Mlog01_M01}(b)}. 
{By contrast, for masses smaller than 1.05 M$_{\odot}$ and larger than 1.3 M$_{\odot}$, the $M_{\rm log}$ relation shows differences in its distribution compared to $M_{\rm sca}$.}
%By contrast, for masses smaller than 1.05 M$_{\odot}$ and larger than 1.3 M$_{\odot}$, the $M_{\rm log}$ relation provides a notable improvement. 
Stars within the 1.05–1.3 M$_{\odot}$ range are generally hot. {In this mass range, the scatters between $M_{\rm log1}$ and $M_1$ reach their maximum, indicating that the differences between the scaling relations are most pronounced for hotter MS stars.
Among these stars, and particularly for hot stars, the $\nu_{\rm min0}$ value shows greater scatter compared to $\nu_{\rm min1}$.
This is due to the fact that $\nu_{\rm min0}$ is relatively shallow in hot stars, leading to higher uncertainties.}
%\textbf{Furthermore, for the masses determined from $\nu_{\rm min0}$, we obtain $r = 0.8757$ and ${\rm RMS} = 0.0752$, while for those derived from $\nu_{\rm min1}$ we find $r = 0.9365$ and ${\rm RMS} = 0.0565$.} These results clearly indicate that the $M_{\rm log,1}$ relation derived from $\nu_{\rm min1}$ is in better agreement than the $M_{\rm log,0}$ relation obtained from $\nu_{\rm min0}$.

The derived masses are expected to exhibit more scatter compared to the radii, due to the contribution of uncertainties in both $g$ and $R$ to the mass uncertainty. %Moreover, despite some scatter, Fig.~\ref{ms_Mlog01_M01} shows good agreement with the results of \cite{2019MNRAS.489.1753Y}. 
The masses presented in Fig.~\ref{ms_Mlog01_M01} correspond to MS stars, in which $\delta\nu_{02}$ is known to have big values. 
{In addition, compared to the study by \cite{2019MNRAS.489.1753Y}, the direct inclusion of $Z$ introduces a methodological difference in the new scaling relation for the mass.}
%{In addition, compared to the study by \cite{2019MNRAS.489.1753Y}, the direct inclusion of $Z$ improved the new scaling relation for the mass.}
%In the study by \cite{2019MNRAS.489.1753Y}, the effects of the $\delta\nu_{02}$ and $\Gamma_1$ were taken into account when deriving the scaling relations and $Z$ correction was subsequently applied. Therefore, the masses obtained in that study exhibit less scatter compared to the masses computed from the $M_{\rm log}$ relation.
\subsubsection{SG \emph{Kepler} targets}
\begin{figure}
  \centering   \includegraphics[width=1.4\linewidth]{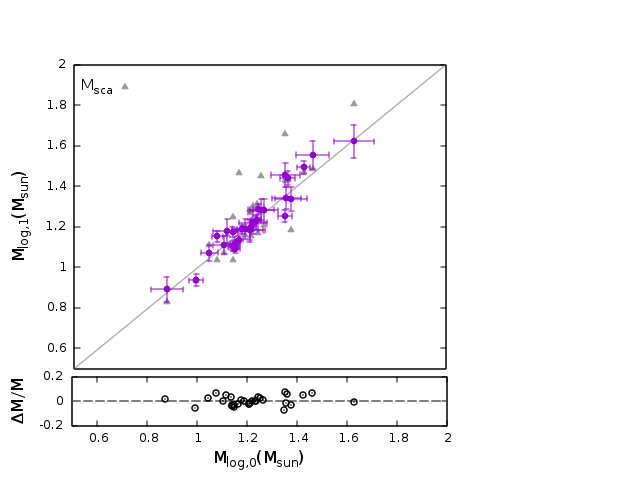}
  \caption{$M_{\rm log,0}$ is plotted with respect to $M_{\rm log,1}$ for SG \emph{Kepler} stars. The triangles indicate the masses determined using the classical scaling relation. {The bottom panel shows the mass differences, which range from 0.07 to -0.07.}}
  \label{pms_Mlog01_M01}
\end{figure}

%In this section, we obtain the scaling relation for $M$ from {\small{MESA}} SG models for the first time and apply it to target stars. We determine the masses $M_{\rm log,0}$ and $M_{\rm log,1}$. The relations between these masses and the target stars' masses are as follows:
The masses $M_{\rm log,0}$ and $M_{\rm log,1}$ are determined, and their correspondence with the target stars' masses is given below:
 \begin{equation}
M_{\rm log,0}=(0.998\pm 0.012){M_{0}},
\label{eq:Rlog0_R0_pms} 
\end{equation}

 \begin{equation}
M_{\rm log,1}=(0.998\pm 0.008){M_{1}}.
\label{eq:Rlog1_R0_ms} 
\end{equation} 
When comparing equation (\ref{eq:Rlog0_R0_pms}) with equation (\ref{eq:Rlog1_R0_ms}), the masses determined from $M_{\rm log,0}$ and $M_{\rm log,1}$ yield nearly identical results. This shows that the asteroseismic masses of SG stars can be determined quite accurately. The relations between $M_{\rm log,1}$ and $M_{\rm log,0}$ is shown in Fig.~\ref{pms_Mlog01_M01}, while $\Delta M / M$  ($(M_{\rm log,1} - M_{\rm log,0}) / M_{\rm log,0}$), is plotted against $M_{\rm log,0}$ in the same figure.
The calculated masses are given in  Table \ref{tab:seismic_parameters_postMS_Kls}. 

The SG models $M_{\rm log,0}$ and $M_{\rm log,1}$ masses agree quite well with each other. However, the masses derived from the scaling relation based on $\nu_{\rm min1}$ provide a slight improvement compared to those determined from $\nu_{\rm min0}$. 
%\textbf{A direct comparison of the two reference-frequency based solutions shows that the $\nu_{\rm min1}$ relation yields smaller residuals and more consistent results throughout the SG mass range, which suggests that $\nu_{\rm min1}$ is a more reliable indicator for evolved stars.} 
This can be explained as follows: these stars are evolved stars that have separated from the MS. As stars evolve, the depth of the $\nu_{\rm min0}$ gradually decreases and eventually becomes undetectable in advanced evolutionary stages. On the other hand, the $\nu_{\rm min1}$ frequency, similar to MS stars, is much deeper compared to other minima in SG stars as well. 

In deriving the mass relations, our analysis yields results consistent with those of \cite{2019MNRAS.489.1753Y}, as we use a similar methodological approach. However, the present study introduces several improvements, such as obtaining simultaneous solutions for $\Delta\nu$, $Z$ and $\nu_{\rm min}$.
{In general, in Fig.~\ref{pms_Mlog01_M01}, the masses determined from the relation $M_{\rm log}$ exhibit less scatter compared to the masses determined from $M_{\rm sca}$. This indicates differences in the distributions obtained from the two scaling relations for post-MS stars. However, comparisons with independent mass and radius measurements are required to determine which scaling relation provides the most accurate estimates of stellar $M$ and $R$.}
%However, a direct assessment of accuracy requires independent constraints.

%In general, in Fig.~\ref{pms_Mlog01_M01}, the masses determined from the relation $M_{\rm log}$ exhibit less scatter and better consistency compared to the masses determined from $M_{\rm sca}$. This result demonstrates that the masses of post-MS stars can also be calculated with high precision using a new relation determined from the reference frequencies, independently of the classical scaling relation.
\subsection{Comparison with detailed seismic modelling}
\begin{figure}
  \centering   \includegraphics[width=1.0\linewidth]{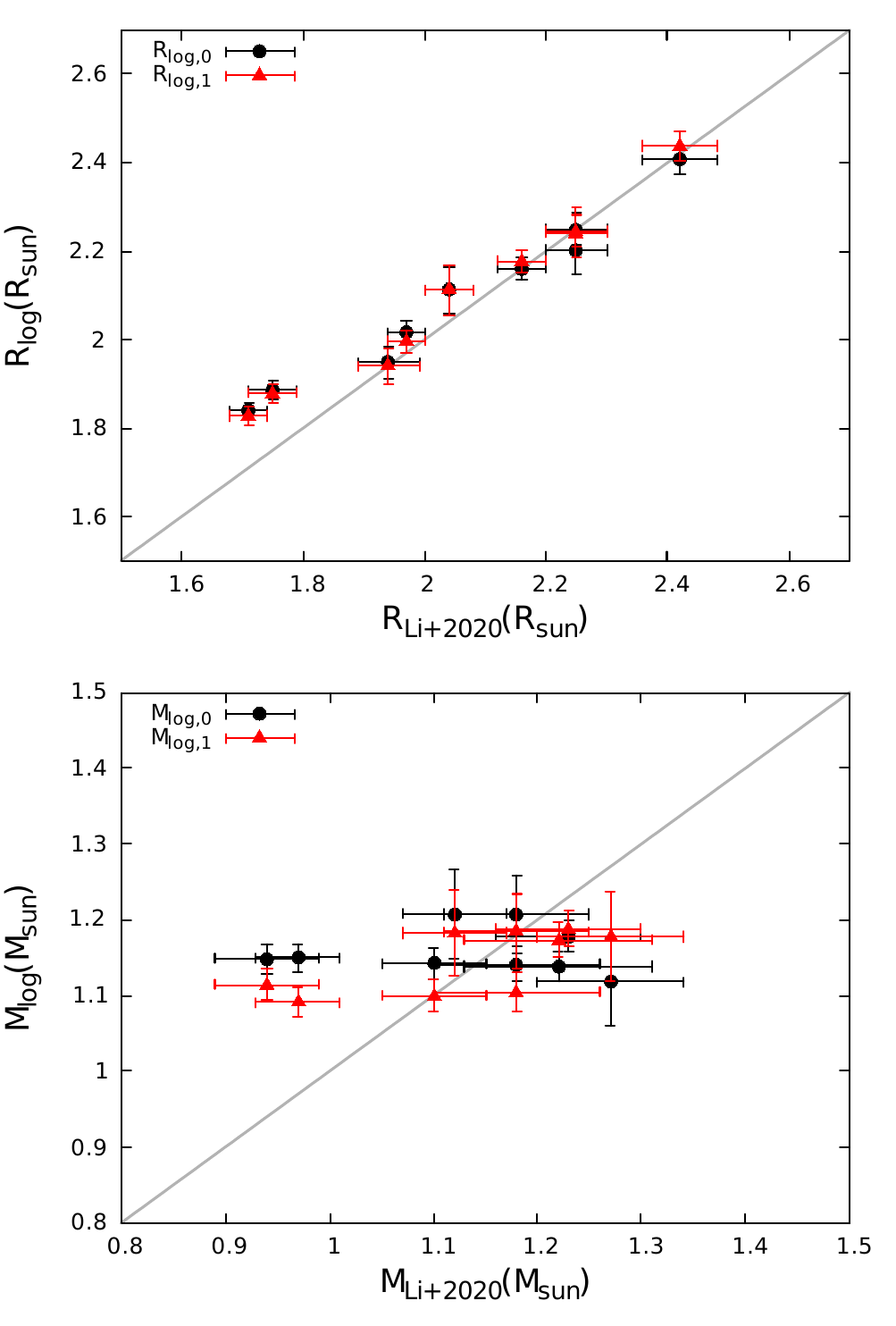}
  \caption{{Comparison of the radii (top panel) and masses (bottom panel) derived in this study with those reported by \citet{2020MNRAS.495.3431L} for the nine SG stars. The circles and triangles represent results obtained using two different minima ($\nu_{\rm min0}$ and $\nu_{\rm min1}$)}.}
  \label{fig:MRlog_krs_Li+2020}
\end{figure}

To more thoroughly evaluate the reliability of the reference frequency-based approach, the results obtained in this study are compared with previous studies in which masses and radii were calculated.
In \cite{2020MNRAS.495.3431L} and \cite{2017ApJ...835..173S}, stellar masses and radii were derived using the detailed asteroseismic modelling approach, in which individual oscillation frequencies are fitted to theoretical models. Of the subgiant stars analysed in \cite{2020MNRAS.495.3431L}, nine are also investigated in the present study. {%Fig. \ref{fig:MRlog_krs_Li+2020} shows a comparison between the masses and radii derived in this study and those reported in that study for these nine stars.The calculated masses are given in  Table \ref{tab:seismic_parameters_postMS_Kls}.
Fig. \ref{fig:MRlog_krs_Li+2020} compares the masses and radii derived here with those reported by \cite{2020MNRAS.495.3431L} for these nine stars, while the calculated masses and radii are listed in Table \ref{tab:seismic_parameters_postMS_Kls}.}

{The radii computed in this study are in good agreement with those reported by \cite{2020MNRAS.495.3431L}; for six stars (KIC 5955122, KIC 7976303, KIC 8702606, KIC 11395018, KIC 11414712, and KIC 12508433), the derived values fall within the specified uncertainty ranges. For two stars (KIC 8524425 and KIC 10920273), the masses and radii lie slightly outside these uncertainties. 
The mass and radius of KIC 8524425 reported by \cite{2020MNRAS.495.3431L} are $0.97\pm0.04$ M$_\odot$ and $1.71\pm0.03$ R$_\odot$, respectively. In this study, the mass and radius derived from $\nu_{\rm min1}$ are $1.09 \pm 0.02$ M$_\odot$ and $1.83\pm 0.02$ R$_\odot$, respectively.
For KIC 10018963, the mass and radius derived using {logarithmic method} based on $\nu_{\rm min1}$ are identical to those obtained in that study. These results indicate that subgiant stellar radii estimated using reference frequencies yield results that are consistent with those obtained from seismic modelling, despite differences in methodology.}

In this study, the mass range obtained for these nine stars using  $\nu_{\rm min0}$ and $\nu_{\rm min1}$ (1.09–1.21 M$_\odot$) is narrower than that reported by \cite{2020MNRAS.495.3431L} (0.94–1.27 M$_\odot$). This difference is likely driven by variations in the adopted observational parameters, particularly the metallicity values used in the scaling relations.
To test this effect, a small sample of stars was re-analysed using the [Fe/H] values taken from \cite{2020MNRAS.495.3431L}, and the resulting masses were found to be in better agreement with those of \cite{2020MNRAS.495.3431L}.

\begin{table*}
  \centering
  \caption{Asteroseismic parameters of SG \emph{Kepler} targets. The KIC ID of each star is given in the first column. {The value of $\nu_{\rm max}$ taken from \protect\cite{2016MNRAS.462.1577Y}}. The parameters $\nu_{\rm min0}$, $\nu_{\rm min1}$, $\nu_{\rm min2}$, $\nu_{\rm min1}$, $\braket{\Delta\nu}$, $\delta\nu_{02}$, $T_{\rm eS}$ (effective temperatures from spectra), {$M_{\rm 0}$}, {$M_{\rm 1}$}, {$R_{\rm 0}$}, and {$R_{\rm 1}$} are from \protect\citet{2019MNRAS.489.1753Y}. {$M_{\rm 0}$}, {$M_{\rm 1}$}, {$R_{\rm 0}$}, and {$R_{\rm 1}$} correspond to {$M_{\rm sis0}$}, {$M_{\rm sis1}$}, {$R_{\rm sis0}$}, and {$R_{\rm sis1}$}, respectively. The fundamental parameters {$M_{\rm log,0}$}, {$M_{\rm log,1}$}, {$R_{\rm log,0}$} and {$R_{\rm log,1}$} are obtained from the {\small {MESA}} models.}  
  \resizebox{\textwidth}{!}{%
    \begin{tabular}{clllllllllllllllll}
    \hline
    \multicolumn{1}{l}{KIC ID} & $\nu_{\rm max}$ & $\nu_{\rm min0}$ & $\nu_{\rm min1}$  & $\nu_{\rm min2}$  &  $\braket{\Delta\nu}$  & $\braket{\delta\nu_{02}}$   & T$_{eS}$  & {$M_{\rm 0}$}    & {$M_{\rm 1}$}    & {$R_{\rm 0}$}    & {$R_{\rm 1}$}   & {$M_{\rm log,0}$}   & {$M_{\rm log,1}$} & {$R_{\rm log,0}$}   & {$R_{\rm log,1}$} \\
          & ($\mu$Hz) & ($\mu$Hz)   & ($\mu$Hz)   & ($\mu$Hz)   & ($\mu$Hz)   & ($\mu$Hz)   & (K)  & (M$_{\odot}$)  & (M$_{\odot}$)  & (R$_{\odot}$)  & (R$_{\odot}$)  & (M$_{\odot}$)  & (M$_{\odot}$)  & (R$_{\odot}$) & (R$_{\odot}$)  \\
    \hline
    \multicolumn{1}{l}{1435467} & 1324  & 1626.4 & 1274  & ---     & 70.9  & 4.8   & 6264  & 1.24  & 1.27  & 1.65  & 1.66  & 1.24  & 1.29  & 1.66  & 1.67\\
          &       &       &       &       &       &       &       &       &       &       &       & 0.02  & 0.02  & 0.02  & 0.02\\
    \multicolumn{1}{l}{3424541} & 745   & 1046.6 & 755.4 & ---     & 41.1  & 4.7   & 6165  & 1.80  & 1.74  & 2.65  & 2.64  & 1.63  & 1.62  & 2.61  & 2.61 \\
          &       &       &       &       &       &       &       &       &       &       &       & 0.08  & 0.08  & 0.08  & 0.08\\
    \multicolumn{1}{l}{3632418} & 1159  & 1370.9 & 1055  & ---     & 60.4  & 3.8   & 6148  & 1.24  & 1.26  & 1.83  & 1.83  & 1.25  & 1.28  & 1.85  & 1.86 \\
          &       &       &       &       &       &       &       &       &       &       &       & 0.05  & 0.05  & 0.03  & 0.04\\
    \multicolumn{1}{l}{5955122} & 861   & 952.7 & 717.9 & ---    & 49.4  & 4.8   & 5952  & 1.20  & 1.21  & 2.12  & 2.13  & 1.21  & 1.18  & 2.11  & 2.11  \\
          &       &       &       &       &       &       &       &       &       &       &       & 0.06  & 0.06  & 0.05  & 0.06 \\
    \multicolumn{1}{l}{6508366} & 926   & 1267.5 & 978.3 & 672.2 & 51.5  & 3.3   & 6354  & 1.43  & 1.46  & 2.11  & 2.12  & 1.42  & 1.49  & 2.14  & 2.16 \\
          &       &       &       &       &       &       &       &       &       &       &       & 0.03  & 0.03  & 0.03  & 0.03\\
    \multicolumn{1}{l}{6603624} & 2402  & 2529.7 & 2080.5 & ---    & 109.7 & 5.5   & 5625  & 0.98  & 1.03  & 1.14  & 1.15  & 1.08  & 1.15  & 1.18  & 1.20 \\
          &       &       &       &       &       &       &       &       &       &       &       & 0.02  & 0.03  & 0.02  & 0.02 \\
    \multicolumn{1}{l}{6679371} & 908   & 1284.8 & 1000.6 & 725.8 & 50.6  & 4.1   & 6344  & 1.51  & 1.55  & 2.17  & 2.18  & 1.46  & 1.55  & 2.18  & 2.21 \\
          &       &       &       &       &       &       &       &       &       &       &       & 0.07  & 0.07  & 0.05  & 0.05 \\
    \multicolumn{1}{l}{6933899} & 1391  & 1538.7 & 1178  & ---    & 71.8  & 4.9   & 5837  & 1.14  & 1.15  & 1.61  & 1.61  & 1.19  & 1.19  & 1.63  & 1.63 \\
          &       &       &       &       &       &       &       &       &       &       &       & 0.05  & 0.05  & 0.03  & 0.04 \\
    \multicolumn{1}{l}{7103006} & 1124  & 1432.9 & 1134.4 & 790.3 & 60.1  & 4.5   & 6394  & 1.36  & 1.41  & 1.90  & 1.91  & 1.36  & 1.44  & 1.91  & 1.93\\
          &       &       &       &       &       &       &       &       &       &       &       & 0.03  & 0.04  & 0.03  & 0.04\\
    \multicolumn{1}{l}{7799349} & 561   & 580.6 & 448.6 & ---    & 33.2  & 3.4   & 4954  & 1.03  & 1.08  & 2.63  & 2.66  & 1.22  & 1.22  & 2.78  & 2.80\\
          &       &       &       &       &       &       &       &       &       &       &       & 0.06  & 0.06  & 0.06  & 0.06\\
    \multicolumn{1}{l}{7976303} & 851   & 1036.3 & 754   & ---    & 51    & 4.5   & 6053  & 1.15  & 1.13  & 2.02  & 2.01  & 1.14  & 1.10  & 2.02  & 2.00\\
          &       &       &       &       &       &       &       &       &       &       &       & 0.02  & 0.02  & 0.02  & 0.02\\
    \multicolumn{1}{l}{8026226} & 545   & 687.7 & 479.1 & ---    & 34.6  & 3.6   & 6230  & 1.39  & 1.32  & 2.80  & 2.78  & 1.35  & 1.25  & 2.78  & 2.74\\
          &       &       &       &       &       &       &       &       &       &       &       & 0.03  & 0.03  & 0.05  & 0.05 \\
    \multicolumn{1}{l}{8228742} & 1171  & 1375.9 & 1036.9 & ---    & 62    & 4.8   & 6042  & 1.24  & 1.23  & 1.81  & 1.81  & 1.23  & 1.23  & 1.81  & 1.81\\
          &       &       &       &       &       &       &       &       &       &       &       & 0.02  & 0.02  & 0.02  & 0.02\\
    \multicolumn{1}{l}{8524425} & 1081  & 1128  & 831.3 & ---    & 59.4  & 5     & 5634  & 1.07  & 1.05  & 1.81  & 1.81  & 1.15  & 1.09  & 1.84  & 1.83\\
          &       &       &       &       &       &       &       &       &       &       &       & 0.02  & 0.02  & 0.02  & 0.02\\
    \multicolumn{1}{l}{8561221} & 491   & 488   & 370.5 & ---    & 29.8  & 2.4   & 5245  & 0.99  & 1.04  & 2.80  & 2.82  & 1.16  & 1.14  & 2.94  & 2.95 \\
          &       &       &       &       &       &       &       &       &       &       &       & 0.02  & 0.02  & 0.02  & 0.02\\
    \multicolumn{1}{l}{8694723} & 1384  & 1661.8 & 1261.7 & ---    & 74.9  & 5.4   & 6258  & 1.10  & 1.09  & 1.53  & 1.53  & 1.10  & 1.11  & 1.54  & 1.53\\
          &       &       &       &       &       &       &       &       &       &       &       & 0.05  & 0.05  & 0.03  & 0.03\\
    \multicolumn{1}{l}{8702606} & 664   & 688.7 & 554.5 & ---    & 39.7  & 3.5   & 5540  & 1.03  & 1.12  & 2.34  & 2.38  & 1.14  & 1.17  & 2.41  & 2.44 \\
          &       &       &       &       &       &       &       &       &       &       &       & 0.02  & 0.02  & 0.03  & 0.03\\
    \multicolumn{1}{l}{8760414} & 2384  & 2628.3 & 2041.6 & ---    & 117.1 & 5.6   & 5850  & 0.85  & 0.85  & 1.03  & 1.03  & 0.87  & 0.89  & 1.05  & 1.04\\
          &       &       &       &       &       &       &       &       &       &       &       & 0.07  & 0.06  & 0.03  & 0.03 \\
    \multicolumn{1}{l}{9574283} & 455   & 459.3 & 370.6 & ---    & 29.9  & 3     & 5120  & 0.90  & 0.99  & 2.72  & 2.77  & 1.04  & 1.07  & 2.83  & 2.87\\
          &       &       &       &       &       &       &       &       &       &       &       & 0.03  & 0.04  & 0.08  & 0.08\\
    \multicolumn{1}{l}{9812850} & 1195  & 1558  & 1171.2 & 877.2 & 65.1  & 4.1   & 6258  & 1.26  & 1.25  & 1.74  & 1.74  & 1.27  & 1.28  & 1.76  & 1.76\\
          &       &       &       &       &       &       &       &       &       &       &       & 0.06  & 0.06  & 0.04  & 0.04\\
    \multicolumn{1}{l}{10018963} & 987   & 1168.7 & 866   & ---    & 55.2  & 5.1   & 6145  & 1.23  & 1.21  & 1.96  & 1.96  & 1.21  & 1.18  & 1.95  & 1.94\\
          &       &       &       &       &       &       &       &       &       &       &       & 0.05  & 0.05  & 0.04  & 0.04 \\
    \multicolumn{1}{l}{10162436} & 968   & 1370.4 & 977   & 671.1 & 55.5  & 3.6   & 6149  & 1.39  & 1.32  & 1.98  & 1.97  & 1.37  & 1.34  & 2.01  & 2.00\\
          &       &       &       &       &       &       &       &       &       &       &       & 0.06  & 0.06  & 0.04  & 0.05\\
    \multicolumn{1}{l}{10355856} & 1330  & 1823.7 & 1308.8 & ---    & 68.1  & 4.7   & 6351  & 1.38  & 1.30  & 1.72  & 1.70  & 1.35  & 1.34  & 1.74  & 1.72\\
          &       &       &       &       &       &       &       &       &       &       &       & 0.06  & 0.06  & 0.03  & 0.04 \\
    \multicolumn{1}{l}{10920273} & 1024  & 1103.1 & 826.6 & ---    & 57.1  & 4.9   & 5710  & 1.09  & 1.09  & 1.86  & 1.86  & 1.15  & 1.11  & 1.88  & 1.88\\
          &       &       &       &       &       &       &       &       &       &       &       & 0.02  & 0.02  & 0.02  & 0.02\\
    \multicolumn{1}{l}{11244118} & 1420  & 1526.8 & 1169.9 & ---    & 71.3  & 5.5   & 5745  & 1.20  & 1.21  & 1.65  & 1.65  & 1.23  & 1.23  & 1.66  & 1.66\\
          &       &       &       &       &       &       &       &       &       &       &       & 0.02  & 0.02  & 0.02  & 0.02\\
    \multicolumn{1}{l}{11253226} & 1638  & 2150.4 & 1684.6 & 1194.6 & 76.9  & 4.4   & 6410  & 1.35  & 1.37  & 1.56  & 1.57  & 1.35  & 1.46  & 1.60  & 1.61\\
          &       &       &       &       &       &       &       &       &       &       &       & 0.06  & 0.06  & 0.03  & 0.04\\
    \multicolumn{1}{l}{11395018} & 834   & 875.3 & 685.2 & ---    & 47.3  & 4.2   & 5445  & 1.06  & 1.11  & 2.10  & 2.12  & 1.18  & 1.19  & 2.16  & 2.18 \\
          &       &       &       &       &       &       &       &       &       &       &       & 0.02  & 0.02  & 0.02  & 0.03\\
    \multicolumn{1}{l}{11414712} & 707   & 781.2 & 586.9 & ---    & 43.9  & 4.1   & 5635  & 1.07  & 1.08  & 2.22  & 2.22  & 1.14  & 1.10  & 2.25  & 2.25\\
          &       &       &       &       &       &       &       &       &       &       &       & 0.02  & 0.03  & 0.04  & 0.04\\
    \multicolumn{1}{l}{11717120} & 585   & 583.7 & 434.3 & ---    & 37.8  & 4.2   & 5150  & 0.86  & 0.88  & 2.32  & 2.32  & 0.99  & 0.94  & 2.38  & 2.36\\
          &       &       &       &       &       &       &       &       &       &       &       & 0.03  & 0.03  & 0.06  & 0.06\\
    \multicolumn{1}{l}{12258514} & 1440  & 1667.1 & 1251.8 & ---    & 74.5  & 4.9   & 5990  & 1.19  & 1.18  & 1.58  & 1.58  & 1.21  & 1.20  & 1.60  & 1.59\\
          &       &       &       &       &       &       &       &       &       &       &       & 0.02  & 0.02  & 0.02  & 0.02\\
    \multicolumn{1}{l}{12508433} & 793   & 786.7 & 650.6 & ---    & 44.9  & 3.8   & 5134  & 0.94  & 1.04  & 2.09  & 2.13  & 1.12  & 1.18  & 2.20  & 2.24\\
          &       &       &       &       &       &       &       &       &       &       &       & 0.06  & 0.06  & 0.05  & 0.06 \\
    \hline
    \end{tabular}%
    }  \label{tab:seismic_parameters_postMS_Kls}
\end{table*}%
\section{Conclusions}
Determining the masses and radii of stars with high precision is crucial for understanding the physical processes occurring in their interiors. By using the oscillation frequencies of solar-like oscillating stars, highly accurate solutions for mass and radius can be obtained. {In this study, we analyse the oscillation frequencies of stellar models obtained using the {\small{MESA}} evolution code and derive new scaling relations for $M$ and $R$ based on reference frequencies.}

{We first derived simultaneous three-parameter solutions for $g$ and $R$ using {\small{MESA}} MS and SG models, based on $\nu_{\rm min}$, $\Delta\nu$, and $Z_{\rm mod}$. The differences between the derived scaling-relation values and the corresponding model values are less than 1 per cent for the radius and 3 per cent for the mass in most models, indicating a high consistency.
%The scatter between the derived scaling-relation values and the corresponding model values is approximately 0.6–0.7\% for the radius and about 2\% for the mass, indicating a high consistency. 
The mass was subsequently calculated using the derived $g$ and $R$ values. In particular, this study presents, for the first time, radius and mass scaling relations based on $\nu_{\rm min}$ for SG models. Furthermore, these relations allow $M$ and $R$ to be reliably determined using only observational parameters, without the need for detailed stellar modelling.}

%The mass and radius scaling relations derived from the {\small MESA} evolution code are applied to the \emph{Kepler} target stars.
The new scaling relations are applied to the \emph{Kepler} target stars. The seismic solutions based on $\nu_{\rm min}$ show excellent agreement with the \emph{Kepler} data. These findings demonstrate that the fundamental parameters of evolved solar-like stars can be reliably determined using the proposed scaling relations. In future research, these scaling relations can be derived for stars with a wider range of masses, radii, and metallicities. 
%By using a fitting formula that includes the $\braket{\delta\nu_{02}}$, which is one of the seismic parameters important in stellar evolution, the masses and radii of stars can be determined more accurately. Moreover, their accuracy can be further evaluated using data from space missions such as TESS and \emph{Kepler}.

{%In this study, we analyzed the oscillation frequencies of the {\small{CESAM}} 1.0 and 1.1 M$_{\odot}$ models, determining $\braket{\Delta\nu}$, $\braket{\delta\nu{02}}$ and the reference frequencies. Using these seismic parameters, we also tested the model dependence of the scaling relations calibrated.

%Oscillation frequencies of the {\small{CESAM}} 1.0 and 1.1 M$_{\odot}$ models are analyzed to determine $\braket{\Delta\nu}$, $\braket{\delta\nu{02}}$, and the reference frequencies. These seismic parameters are used to test the model dependence of the calibrated scaling relations. The {\small MESA}-calibrated scaling relations are applied to the {\small CESAM} models, and the resulting radii and masses are directly compared with the corresponding model values. The results indicate that, within the explored mass range, the scaling relations do not exhibit a significant dependence on the adopted stellar evolution code.}

{In addition, we acknowledge that surface effects associated with the modelling of near-surface layers can influence the oscillation frequencies used in this study. Although the application of explicit surface effect corrections is beyond the scope of this study, such effects may contribute to systematic uncertainties in the derived stellar parameters. Future studies incorporating surface-effect corrections are expected to improve the derived stellar mass and radius estimates.}

\section*{Acknowledgements}

This work is supported by the Scientific and Technological Research Council of Turkey (TÜBİTAK: 123F019). This study is part of the author's MSc thesis submitted to Ege University.

\section*{Data Availability}

The data underlying this article will be shared on reasonable request to the corresponding author. 
%%%%%%%%%%%%%%%%%%%%%%%%%%%%%%%%%%%%%%%%%%%%%%%%%%
%\section*{Data Availability}

%The inclusion of a Data Availability Statement is a requirement for articles published in MNRAS. Data Availability Statements provide a standardised format for readers to understand the availability of data underlying the research results described in the article. The statement may refer to original data generated in the course of the study or to third-party data analysed in the article. The statement should describe and provide means of access, where possible, by linking to the data or providing the required accession numbers for the relevant databases or DOIs.



\bibliographystyle{mnras}

%%%%%%%%%%%%%%%%%%%%%%%%%%%%%%%%%%%%%%%%%%%%%%%%%%

%%%%%%%%%%%%%%%%% APPENDICES %%%%%%%%%%%%%%%%%%%%%
\appendix
\section{Test of Model Dependence Using {\small CESAM} Models} \label{App_A}
{To explicitly test the model dependence of the calibrated scaling relations, we applied the {\small MESA}-calibrated relations to the asteroseismic parameters of the {\small{CESAM}} models. The {\small{CESAM}} seismic parameters computed in this study and used in the scaling relations are listed in Tables \ref{tab:CESAM_ak} and \ref{tab:CESAM_aks}. The resulting radii and masses were directly compared with the true {\small{CESAM}} model values, as shown in Figs \ref{fig:CESAM_with_MESA_Rlog} and \ref{fig:CESAM_with_MESA_Mlog}.}

{Regarding the radius relations, the radii obtained using both $\nu_{\rm min0}$ ($R_{\rm log,0}$) and $\nu_{\rm min1}$ ($R_{\rm log,1}$) are systematically slightly larger than the {\small{CESAM}} model radii. When all models are considered together, the mean radius difference is approximately 2 per cent. The residual panels show that the differences remain nearly constant across the examined radius range, thus indicating no significant radius-dependent trend. The radius comparisons clearly show improved agreement for the 1.1 M$_\odot$ models. As seen in Fig. \ref{fig:CESAM_with_MESA_Rlog}(a), the mean radius difference is 1.26 per cent for the 1.1 M$_\odot$ models, while it reaches 3.05 per cent for the 1.0 M$_\odot$ models. 
%The internal scatter is lower for the 1.1 M$_\odot$ models(0.53\% vs. 0.74\%). 
This indicates that the radius scaling relation performs more consistently and appears better calibrated in the 1.1 M$_\odot$ mass regime. A similar trend is observed in Fig. \ref{fig:CESAM_with_MESA_Rlog}(b) for the $R_{\rm log,1}$ relation, where the 1.1 M$_\odot$ models show smaller systematic offsets and lower dispersion.}
\begin{figure}
  \centering    \includegraphics[width=1.4\linewidth]{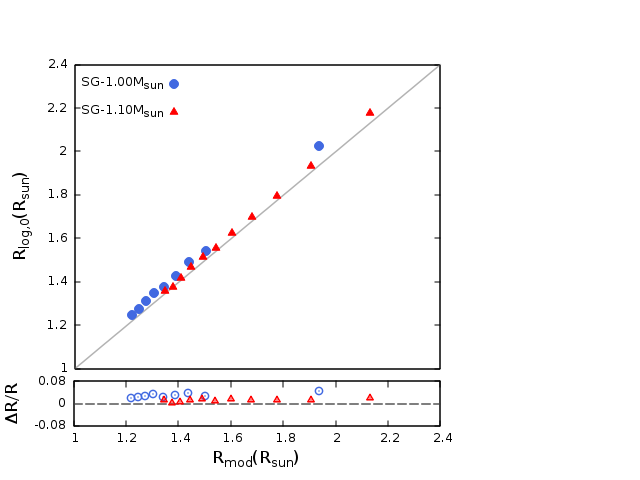}    \includegraphics[width=1.4\linewidth]{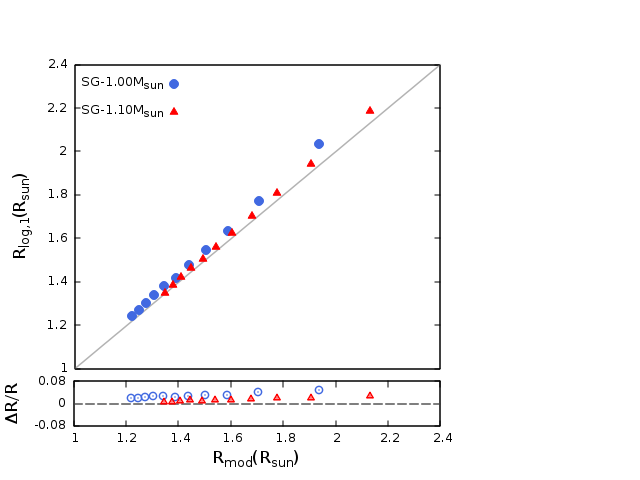}    
\caption{{{\small{CESAM}} $R_{\rm log}$ radii obtained using equation~(\ref{eq:ms_Msis_R}) for SG models, plotted against the {\small{CESAM}} model radius $R_{\rm mod}$. (a) Relation between $R_{\rm log,0}$ and $R_{\rm mod}$. (b) Relation between $R_{\rm log,1}$ and $R_{\rm mod}$. Circles represent 1.0 M$_\odot$ {\small{CESAM}} models, while filled triangles denote 1.1 M$_\odot$ {\small{CESAM}} models. The lower panels show the variation of the residuals ($\Delta R / R$) with $R_{\rm mod}$. In the bottom panels, the same symbols as in the upper panels are shown with open markers to represent the residuals.}}
  \label{fig:CESAM_with_MESA_Rlog}
\end{figure}

{For the masses, the agreement was examined considering the limited mass range of the {\small CESAM} models (1.0–1.1 M$_\odot$). When all models are considered together, the mean mass difference is $\sim$ 7 per cent, with an internal scatter of 1-3 per cent.
There is a significant difference compared to the $\sim$ 3 per cent mass discrepancy obtained when using models generated with the same evolutionary code. This suggests that the discrepancy may arise from hidden dependencies on the physical assumptions adopted in the construction of the stellar models used to calibrate the scaling relations. Similar behaviour has been reported by \cite{2025A&A...699A.325V}, especially for mass estimates in model-based corrections to classical scaling relations.}

%Similar model-dependent behaviour has been reported in the context of model-based corrections to classical scaling relations (e.g. Valle et al. 2025).}
%{For the masses, considering the limited mass range of the {\small CESAM} models (1.0–1.1 M$_\odot$), the overall agreement can be reasonable. When all models are considered together, the mean mass difference is approximately 7\%, with an internal scatter in the range of 1-3\%. This behaviour is physically expected, since the mass scaling relation depends quadratically on the radius ($M \propto g R^{2}$), meaning that even small radius differences are amplified in the derived masses. 

%For the 1.1 M$_\odot$ models, the masses derived from $\nu_{\rm min1}$ provide more consistent mass estimates, with a scatter of less than 1\%.Although a small systematic offset is observed for the 1.0 M$_\odot$ models, no strong mass-dependent trend is detected within the explored mass interval.
\begin{figure}
  \centering   \includegraphics[width=1.4\linewidth]{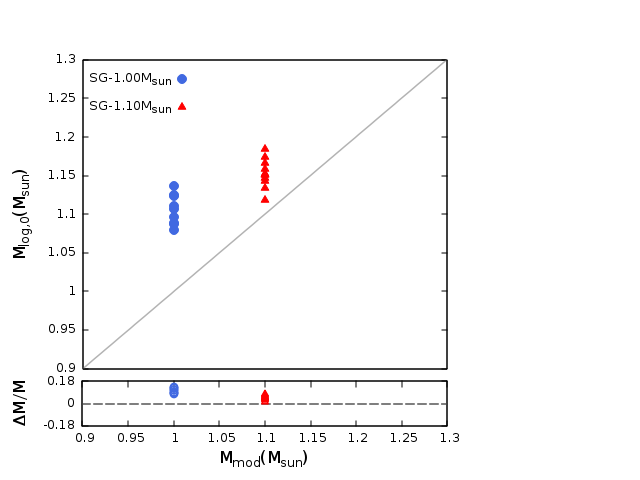}    \includegraphics[width=1.4\linewidth]{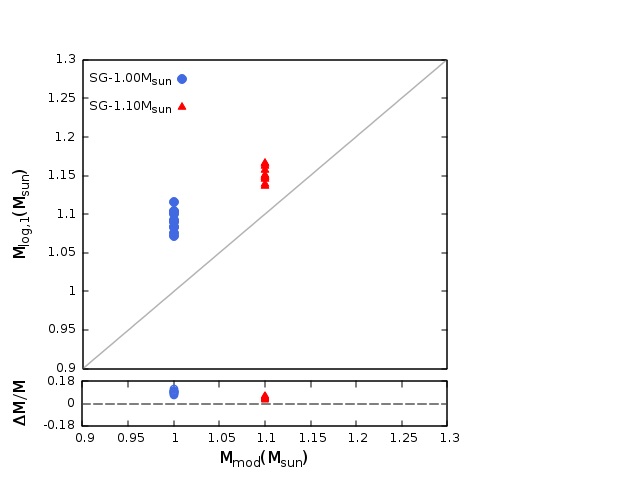}  
  \caption{{Masses computed using {\small{CESAM}} $M_{\rm log}$ radii, obtained from scaling relations calibrated with {\small{MESA}} models, plotted against the {\small{CESAM}} model masses. (a) $M_{\rm log,0}$ versus $M_{\rm mod}$. (b) $M_{\rm log,1}$ versus $M_{\rm mod}$. Circles represent 1.0 M$_\odot$ {\small{CESAM}} models, while filled triangles denote 1.1 M$_\odot$ {\small{CESAM}} models. The bottom panels show the variation of the residuals ($\Delta M / M$) with $M_{\rm mod}$, and the same symbols as in the upper panels are shown with open markers to represent the residuals.}}
  \label{fig:CESAM_with_MESA_Mlog}
\end{figure}

{In general, these results indicate that the scaling relations calibrated using the {\small MESA} models can be successfully applied to independent {\small CESAM} models. Within the explored range, the radius estimates show no strong dependence on the adopted stellar evolution code, whereas the mass estimates exhibit larger systematic differences. This suggests that the model dependence of the masses should be investigated in more detail in future studies.}

\begin{table*}
  \centering
  \caption{Seismic parameters for {\small{CESAM}} MS models. Effective temperatures ($T_{\rm eff}$) and radii are taken from  \citep{2008Ap&SS.316..173M}.}
    \begin{tabular}{rrrrrrrrrrrSS}
    \hline
    \multicolumn{1}{l}{$M_{\rm mod}$} & \multicolumn{1}{l}{$R_{\rm mod}$} & \multicolumn{1}{l}{$T_{\rm eff}$} & \multicolumn{1}{l}{$\nu_{\rm max}$} & \multicolumn{1}{l}{$\nu_{\rm min0}$} & \multicolumn{1}{l}{$\nu_{\rm min1}$} & \multicolumn{1}{l}{$\nu_{\rm min2}$} & \multicolumn{1}{l}{$\braket{\Delta\nu}$} & \multicolumn{1}{l}{$\braket{\delta\nu_{02}}$} \\
      \multicolumn{1}{l}{(M$_{\odot}$)} & \multicolumn{1}{l}{(R$_{\odot}$)} & \multicolumn{1}{l}{(K)} & \multicolumn{1}{l}{($\mu$Hz)} & \multicolumn{1}{l}{($\mu$Hz)} & \multicolumn{1}{l}{($\mu$Hz)} & \multicolumn{1}{l}{($\mu$Hz)} & \multicolumn{1}{l}{($\mu$Hz)} & \multicolumn{1}{l}{($\mu$Hz)} \\
    \hline
    1     & 0.919 & 5639  & 3701.89 & \multicolumn{1}{c}{---} & 2970.12 & 2288.65 & 153.87 & 13.99 \\
    1     & 0.933 & 5657  & 3583.45 & \multicolumn{1}{c}{---} & 2882.88 & 2219.10 & 150.47 & 13.14 \\
    1     & 0.948 & 5678  & 3463.57 & \multicolumn{1}{c}{---} & 2797.63 & 2146.70 & 146.96 & 12.27 \\
    1     & 0.965 & 5696  & 3341.56 & \multicolumn{1}{c}{---} & 2687.18 & 1957.19 & 143.53 & 11.25 \\
    1     & 0.983 & 5718  & 3213.28 & \multicolumn{1}{c}{---} & 2610.03 & 1900.86 & 139.57 & 10.48 \\
    1     & 1.003 & 5738  & 3080.43 & 3216.26 & 2528.83 & 1840.59 & 135.56 & 9.57\\
    1     & 1.026 & 5756  & 2941.01 & 3118.55 & 2442.08 & 1775.73 & 131.27 & 8.66 \\
    1     & 1.051 & 5771  & 2795.27 & 3014.27 & 2349.75 & 1706.00 & 126.25 & 8.00 \\
    1     & 1.080 & 5780  & 2645.62 & 2909.33 & 2254.25 & 1632.80 & 121.42 & 7.09 \\
    1.1   & 1.036 & 5952  & 3119.08 & 3484.69 & 2632.81 & 1925.58 & 134.98 & 13.20 \\
    1.1   & 1.052 & 5965  & 3020.72 & 3389.31 & 2555.22 & 1873.90 & 131.98 & 12.49 \\
    1.1   & 1.070 & 5981  & 2919.09 & 3282.84 & 2475.76 & 1819.38 & 128.87 & 11.76 \\
    1.1   & 1.088 & 5994  & 2820.22 & 3184.29 & 2399.28 & 1767.36 & 125.79 & 11.04 \\
    1.1   & 1.108 & 6005  & 2716.06 & 3084.00 & 2318.84 & 1712.42 & 122.36 & 10.39 \\
    1.1   & 1.129 & 6015  & 2610.52 & 2980.46 & 2232.76 & 1656.95 & 118.76 & 9.72 \\
    1.1   & 1.153 & 6019  & 2503.88 & 2879.61 & 2159.75 & 1600.56 & 115.25 & 9.01 \\
    1.1   & 1.179 & 6021  & 2394.22 & \multicolumn{1}{c}{---} & 2085.89 & 1542.73 & 111.44 & 8.44 \\
    1.1   & 1.207 & 6020  & 2283.05 & 2608.61 & 2010.55 & 1483.06 & 107.67 & 7.77 \\
    1.1   & 1.238 & 6011  & 2172.44 & 2446.24 & 1909.47 & 1420.66 & 103.85 & 7.10 \\
    1.1   & 1.275 & 5996  & 2052.19 & 2378.31 & 1795.53 & 1281.77 & 99.28 & 6.72 \\
    1.1   & 1.319 & 5987  & 1917.19 & 2280.16 & 1653.78 & 1193.45 & 94.462 & 6.25 \\
    \hline
    \end{tabular}%
  \label{tab:CESAM_ak}%
\end{table*}%

% Table generated by Excel2LaTeX from sheet 'CESAM-postms-tablo'
\begin{table*}
  \centering
  \caption{Seismic parameters for {\small{CESAM}} SG models. Effective temperatures ($T_{\rm eff}$) and radii are taken from  \citep{2008Ap&SS.316..173M}.}
    \begin{tabular}{rrrrrrrrrrrSS}
    \hline
    \multicolumn{1}{l}{$M_{\rm mod}$} & \multicolumn{1}{l}{$R_{\rm mod}$} &
      \multicolumn{1}{l}{$T_{\rm eff}$} & \multicolumn{1}{l}{$\nu_{\rm max}$} & \multicolumn{1}{l}{$\nu_{\rm min0}$} & \multicolumn{1}{l}{$\nu_{\rm min1}$} & \multicolumn{1}{l}{$\nu_{\rm min2}$} & \multicolumn{1}{l}{$\braket{\Delta\nu}$} & 
    \multicolumn{1}{l}{$\braket{\delta\nu_{02}}$} \\
    \multicolumn{1}{l}{(M$_{\odot}$)} & \multicolumn{1}{l}{(R$_{\odot}$)} & \multicolumn{1}{l}{(K)} & \multicolumn{1}{l}{($\mu$Hz)} & \multicolumn{1}{l}{($\mu$Hz)} & \multicolumn{1}{l}{($\mu$Hz)} & \multicolumn{1}{l}{($\mu$Hz)} & \multicolumn{1}{l}{($\mu$Hz)} & \multicolumn{1}{l}{($\mu$Hz)} \\
    \hline 
    1     & 1.219 & 3761  & 2080.63 & 2263.97 & 1731.75 & 1281.16 & 101.58 & 5.40 \\
    1     & 1.245 & 3761  & 1996.78 & 2195.60 & 1676.07 & 1194.09 & 98.60 & 5.22 \\
    1     & 1.272 & 3760  & 1911.32 & 2158.04 & 1617.77 & 1140.86 & 95.54 & 5.19 \\
    1     & 1.304 & 3759  & 1821.53 & 2085.61 & 1552.80 & 1085.86 & 92.12 & 5.35 \\
    1     & 1.342 & 3758  & 1723.17 & 1870.37 & 1452.82 & 1056.56 & 88.24 & 5.48 \\
    1     & 1.385 & 3756  & 1620.24 & 1797.49 & 1337.31 & 1003.80 & 84.32 & 5.51 \\
    1     & 1.347 & 3753  & 1511.04 & 1727.48 & 1265.06 & 939.45 & 79.92 & \multicolumn{1}{c}{---}\\
    1     & 1.501 & 3749  & 1391.67 & 1505.90 & 1179.50 & 823.78 & 75.16 & \multicolumn{1}{c}{---}\\
    1     & 1.584 & 3743  & 1258.23 & \multicolumn{1}{c}{---}& 1030.62 & 756.80 & 69.16 & \multicolumn{1}{c}{---}\\
    1     & 1.703 & 3730  & 1105.28 & \multicolumn{1}{c}{---}& 921.49 & 660.93 & 62.25 & \multicolumn{1}{c}{---}\\
    1     & 1.936 & 3699  & 885.21 & 898.29 & 693.59 & 505.59 & 51.07 & \multicolumn{1}{c}{---}\\
    1     & 2.559 & 3679  & 519.01 & \multicolumn{1}{c}{---}& \multicolumn{1}{c}{---}& \multicolumn{1}{c}{---}& 33.07 & \multicolumn{1}{c}{---} \\
    1.1   & 1.344 & 3777  & 1848.39 & 2168.22 & 1604.65 & 1159.27 & 91.99 & \multicolumn{1}{c}{---}\\
    1.1   & 1.374 & 3776  & 1769.45 & 1973.63 & 1549.12 & 1120.13 & 89.12 & \multicolumn{1}{c}{---}\\
    1.1   & 1.407 & 3775  & 1688.74 & 1913.96 & 1493.45 & 1079.78 & 86.03 & \multicolumn{1}{c}{---}\\
    1.1   & 1.445 & 3774  & 1603.60 & 1878.31 & 1423.66 & 1040.63 & 82.67 & \multicolumn{1}{c}{---}\\
    1.1   & 1.489 & 3772  & 1514.31 & 1806.59 & 1322.30 & 967.83 & 79.25 & \multicolumn{1}{c}{---}\\
    1.1   & 1.540 & 3770  & 1419.17 & 1603.78 & 1237.42 & 884.39 & 75.41 & \multicolumn{1}{c}{---}\\
    1.1   & 1.601 & 3767  & 1317.32 & 1536.02 & 1161.30 & 824.60 & 71.34 & \multicolumn{1}{c}{---}\\
    1.1   & 1.676 & 3762  & 1207.94 & 1356.28 & 1043.59 & 768.31 & 66.53 & \multicolumn{1}{c}{---}\\
    1.1   & 1.775 & 3755  & 1086.75 & 1198.18 & 945.84 & 664.38 & 61.28 & \multicolumn{1}{c}{---}\\
    1.1   & 1.905 & 3741  & 959.19 & 1038.23 & 806.77 & 590.81 & 55.25 & \multicolumn{1}{c}{---}\\
    1.1   & 2.130 & 3705  & 799.67 & 815.12 & 632.48 & 460.59 & 46.55 & \multicolumn{1}{c}{---}\\
    \hline
    \end{tabular}%
  \label{tab:CESAM_aks}%
\end{table*}%
\section{Metallicity Sensitivity of the Radius Relation} \label{App_B}
{To evaluate the effect of metallicity on the derived scaling relation, an alternative scaling relation for the radius-based on MESA SG models and excluding the $Z$ term-has also been derived.  This relation has been obtained only for the reference frequency $\nu_{\rm min1}$ as an example. 
The calculated $R_{\rm log,1}$ radii for SG stars are plotted against the model radius $R_{\rm mod}$ in Fig. \ref{fig:MESA_Rlog1_Zli_Zsiz}. The inclusion of the $Z$ term leads to only a limited change in the radius estimates at the level of the uncertainties. The difference between the $Z$-included and $Z$-excluded radii is less than 2 per cent.}
\begin{figure}
  \centering    \includegraphics[width=1.4\linewidth]{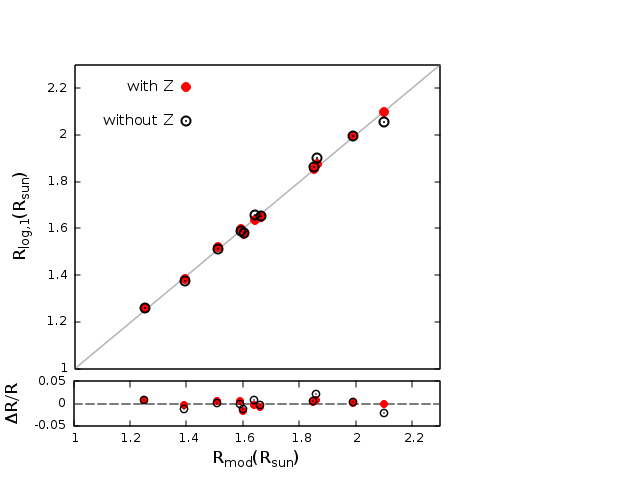}      
\caption{{{\small{MESA}} $R_{\rm log,1}$ radii obtained for SG models, plotted against the {\small{MESA}} model radius $R_{\rm mod}$. Filled circles represent the radii calculated from equation (\ref{eq:ms_Msis_R}), while open circles denote the radii derived from an alternative scaling relation that does not include the $Z$. The bottom panels show the variation of the residuals ($\Delta R / R$) with $R_{\rm mod}$. The same symbols as in the upper panels are used with open markers to represent the residuals. The differences range from 0.02 to -0.02.}}
  \label{fig:MESA_Rlog1_Zli_Zsiz}
\end{figure}
%$R_{\rm q1}$ versus $R_{1}$ is as in Fig.~\ref{Kepler_pms_R01}b. The radii are well matched for the specified mass range. The small difference after 2.4 R$_{\odot}$ in Fig~\ref{Kepler_pms_R01}a decreases here. This shows that the scaling relation for the radius derived from the {\small{CESAM}} SG models can be determined much more accurately for $\nu_{\rm min1}$ than for $\nu_{\rm min0}$.

%%%%%%%%%%%%%%%%%%%%%%%%%%%%%%%%%%%%%%%%%%%%%%%%%%

% Don't change these lines
\bsp	% typesetting comment
\label{lastpage}
\end{document}